\documentclass{aa}  
 \usepackage{caption}
 \usepackage{graphicx}
\usepackage{txfonts}
\usepackage{threeparttable}
\usepackage{CJK}
\begin{document}

   \title{The HI/OH/Recombination line survey of the inner Milky Way (THOR): data release 2 and \ion{H}{i} overview}
        \titlerunning{THOR data release 2}
%(王渊)
\author{
        Y. Wang\inst{1}
        \and
        H. Beuther\inst{1}       
        \and
        M. R. Rugel\inst{2}
        \and
        J. D. Soler\inst{1}
        \and
        J. M. Stil\inst{3}
        \and
        J. Ott\inst{4}
        \and
        S. Bihr\inst{1}
        \and
        N. M. McClure-Griffiths\inst{5}
        \and
        L. D. Anderson\inst{6,7,8}
        \and
        R. S. Klessen\inst{9,10}
        \and
         P. F. Goldsmith\inst{11}
           \and
          N. Roy\inst{12}
          \and
          S. C. O. Glover\inst{9}
          \and
          J. S. Urquhart\inst{13}
          \and
          M. Heyer\inst{14}
          \and
           H. Linz\inst{1}
          \and
          R. J. Smith\inst{15}
          \and
          F. Bigiel\inst{16}
          \and
          J. Dempsey\inst{5}
        \and
          T. Henning\inst{1}
}
\institute{
Max Planck Institute for Astronomy, K\"onigstuhl 17, 69117 Heidelberg, Germany\\
\email{wang@mpia.de}
\and
Max-Planck-Institut f\"ur Radioastronomie, Auf dem H\"ugel 69, 53121 Bonn, Germany
\and
Department of Physics and Astronomy, The University of Calgary, 2500 University Drive NW, Calgary AB T2N 1N4, Canada
\and
National Radio Astronomy Observatory, P.O. Box O, 1003 Lopezville Road, Socorro, NM 87801, USA
\and
Research School of Astronomy and Astrophysics, The Australian National University, Canberra, ACT, Australia
\and
Department of Physics and Astronomy, West Virginia University, Morgantown, WV 26506, USA
\and
Adjunct Astronomer at the Green Bank Observatory, P.O. Box 2, Green Bank WV 24944, USA
\and
Center for Gravitational Waves and Cosmology, West Virginia University, Chestnut Ridge Research Building, Morgantown, WV 26505, USA
\and
Universit\"at Heidelberg, Zentrum f\"ur Astronomie, Institut f\"ur Theoretische Astrophysik, Albert-Ueberle-Str. 2, 69120 Heidelberg, Germany
 \and
Universit\"at Heidelberg, Interdisziplin\"ares Zentrum f\"ur Wissenschaftliches Rechnen, INF 205, 69120, Heidelberg, Germany
 \and
Jet Propulsion Laboratory, California Institute of Technology, 4800 Oak Grove Drive, Pasadena, CA 91109, USA
\and
Department of Physics, Indian Institute of Science, Bengaluru 560012, India
\and
 School of Physical Sciences, University of Kent, Ingram Building, Canterbury, Kent CT2 7NH, UK
 \and
 Department of Astronomy, University of Massachusetts, Amherst, MA01003, USA
\and 
Jodrell Bank Centre for Astrophysics, School of Physics and Astronomy, The University of Manchester, Oxford Road, Manchester, M13 9PL, UK 
\and
Argelander Institut f\"ur Astronomie, Auf dem H\"ugel 71, 53121 Bonn, Germany
}
\date{Received dd, mm, yyyy; accepted dd, mm, yyyy}

% \abstract{}{}{}{}{} 
% 5 {} token are mandatory
 
  \abstract
 {The Galactic plane has been observed extensively by a large number of Galactic plane surveys from infrared to radio wavelengths at an angular resolution below 40\arcsec. However, a 21~cm line and continuum survey with comparable spatial resolution is lacking. }
{The first half of THOR data ($l=14.0\degr-37.9\degr$, and $l=47.1\degr-51.2\degr$, $\lvert b \rvert \leq 1.25\degr$) has been published in our data release 1 paper. With this data release 2 paper, we publish all the remaining spectral line data and Stokes I continuum data with high angular resolution (10\arcsec--40\arcsec), including a new \ion{H}{i} dataset for the whole THOR survey region ($l=14.0-67.4\degr$ and $\lvert b \rvert \leq 1.25\degr$). As we published the results of OH lines and continuum emission elsewhere, we concentrate on the \ion{H}{i} analysis in this paper.}
{With the {\it Karl G. Jansky} Very Large Array (VLA) in C-configuration,  we observed a large portion of the first Galactic quadrant, achieving an angular resolution of $\leq 40$\arcsec. At $L$ Band, the WIDAR correlator at the VLA was set to cover the 21~cm \ion{H}{i} line, four OH transitions, a series of H$n\alpha$ radio recombination lines (RRLs; $n=151$ to 186), and eight 128~MHz-wide continuum spectral windows (SPWs), simultaneously.}
{We publish all OH and RRL data from the C-configuration observations, and a new \ion{H}{i} dataset combining VLA C+D+GBT  (VLA D-configuration and GBT data are from the VLA Galactic Plane Survey) for the whole survey. The \ion{H}{i} emission shows clear filamentary substructures at negative velocities with low velocity crowding. The emission at positive velocities is more smeared-out, likely due to higher spatial and velocity crowding of structures at the positive velocities. Compared to the spiral arm model of the Milky Way, the atomic gas follows the Sagittarius and Perseus Arm well, but with significant material in the inter-arm regions. With the C-configuration-only \ion{H}{i}+continuum data, we produced a \ion{H}{i} optical depth map of the THOR areal coverage from 228 absorption spectra with the nearest-neighbor method. With this $\tau$ map, we corrected the \ion{H}{i} emission for optical depth, and the derived column density is 38\% higher than the column density with optically thin assumption. The total \ion{H}{i} mass with optical depth correction in the survey region is  4.7$\times10^8~M_\odot$, 31\% more than the mass derived assuming the emission is optically thin. If we applied this 31\% correction to the whole Milky Way, the total atomic gas mass would be 9.4--10.5$\times 10^9~M_\odot$. Comparing the \ion{H}{I} with existing CO data, we find a significant increase in the atomic-to-molecular gas ratio from the spiral arms to the inter-arm regions.}%
  % conclusions heading (optional), leave it empty if necessary   {}
{The high-sensitivity and resolution THOR \ion{H}{i} dataset provides an important new window on the physical and kinematic properties of gas in the inner Galaxy. Although the optical depth we derive is a lower limit, our study shows that the optical depth correction is significant for \ion{H}{i} column density and mass estimation. Together with the OH, RRL and continuum emission from the THOR survey, these new \ion{H}{i} data provide the basis for high-angular-resolution studies of the interstellar medium (ISM) in different phases.}

 %\LEt{xxx}
 \maketitle
%
%-------------------------------------------------------------------
\section{Introduction}
The Galactic plane has been observed extensively over the past decades by different survey projects at multiple wavelengths in both continuum and spectral lines, from near infrared (e.g., UKIDSS\footnote{UKIRT Infrared Deep Sky Survey}, \citealt{lucas2008}; {\it Spitzer}/GLIMPSE\footnote{Galactic Legacy Infrared Midplane Survey Extraordinaire }, \citealt{benjamin2003,churchwell2009}, {\it Spitzer}/MIPSGAL\footnote{A 24 and 70 Micron Survey of the Inner Galactic Disk with MIPS}, \citealt{carey2009}, {\it Herschel}/Hi-GAL\footnote{{\it Herschel} Infrared GALactic plane survey}, \citealt{Molinari2010}), to (sub)mm (e.g., ATLASGAL\footnote{APEX Telescope Large Area Survey of the Galaxy}, BGPS\footnote{Bolocam Galactic Plane Survey}, GRS\footnote{The Boston University-Five College Radio Astronomy Observatory Galactic Ring Survey}, MALT90\footnote{The Millimeter Astronomy Legacy Team Survey at 90 GHz}, MALT-45\footnote{The Millimetre Astronomer’s Legacy Team - 45 GHz}, FUGIN\footnote{FOREST unbiased Galactic plane imaging survey with the Nobeyama 45 m telescope}, MWISP\footnote{The Milky Way Imaging Scroll Painting}, \citealt{schuller2009, rosolowsky2010, aguirre2011, csengeri2014, Jackson2006, Foster2011, Jordan2015, Umemoto2017, Su2019}), and radio wavelengths (e.g. MAGPIS\footnote{Multi-Array Galactic Plane Imaging Survey}, CORNISH\footnote{the Co-Ordinated Radio `N' Infrared Survey for High-mass star formation}, CGPS\footnote{The Canadian Galactic Plane Survey}, SGPS\footnote{The Southern Galactic Plane Survey}, VGPS\footnote{The VLA Galactic Plane Survey}, HOPS\footnote{The H$_2$O Southern Galactic Plane Survey}, Sino-German 6~cm survey, \citealt{helfand2006, hoare2012, Taylor2003, McClure2005, stil2006, Walsh2011, Sun2007}). These surveys provide vital data to study and understand the interstellar medium (ISM) in different phases: atomic, molecular, ionized gas, and dust. While many of the surveys have a high angular resolution ($\leq$~40$\arcsec$), the highest angular resolution 21~cm \ion{H}{i} line survey of the northern Galactic plane, VGPS \citep{stil2006}, has a resolution of only 60\arcsec, which makes it difficult to compare with the aforementioned surveys to study the phase transitions of the ISM. 

We therefore initiated the \ion{H}{i}, OH, recombination line survey of the Milky Way (THOR\footnote{\url {http://www.mpia.de/thor/Overview.html}}; \citealt{beuther2016}). A large fraction of the Galactic plane in the first quadrant of the Milky Way ($l=14.0-67.4^\circ$ and $\lvert b \rvert \leq 1.25^\circ$) was observed with the {\it Karl G. Jansky} Very Large Array (VLA) in C-configuration. At $L$ Band, the WIDAR correlator at the VLA was set to cover the 21~cm \ion{H}{i} line, four OH transitions, a series of H$n\alpha$ radio recombination lines (RRLs; $n=151$ to 186), as well as eight 128~MHz wide continuum spectral windows (SPWs), simultaneously. With the C-configuration, we achieve an angular resolution of $<25$\arcsec\ to compare with existing surveys at a matching resolution. The main survey description and data release 1 ($l=14.0\degr-37.9\degr$, and $l=47.1\degr-51.2\degr$) is presented in \citet{beuther2016}. Here, we publish the remaining \ion{H}{i}, OH, and RRL data, including a whole new set of \ion{H}{i} data that combines the existing D-configuration and Green Bank Telescope (GBT) observations to recover the larger scale emission \citep{stil2006}. Scientifically, we focus on an overview of the new \ion{H}{i} data in this paper. The continuum emission, OH absorption and masers from the survey were studied and presented in \citet{Bihr2016}, \citet{Walsh2016}, \citet{rugel2018}, \citet{Wang2018}, and \citet{Beuther2019}. Additionally, \citet{anderson2017} identified 76 new Galactic supernova remnant (SNR) candidates with the continuum data. Using the THOR RRL data, \citet{Rugel2019} studied the feedback in W49A and suggest that star formation in W49A is potentially regulated by feedback-driven and re-collapsing shells. 

Since the 1950s, the Galactic 21~cm \ion{H}{i} line has been extensively observed both in emission and absorption \citep[e.g.,][]{Ewen1951, Muller1951, Heeschen1954, Radhakrishnan1972, Dickey1983, Dickey1990, Gibson2000, Heiles2003a, Li2003, Goldsmith2005}. The atomic gas traced by \ion{H}{i} is widely distributed in the Galaxy \citep[e.g.,][]{Dickey1990, Hartmann1997, Kalberla2009}, and numerous \ion{H}{i} surveys have been carried out \citep[e.g.,][]{Kalberla2005, Stanimirovic2006, stil2006, McClure-Griffiths2009, Peek2011, Dickey2013, Winkel2016, Peek2018} to study the properties of atomic gas and the Galactic spiral structure \citep[e.g.,][]{vandeHulst1954, Oort1958, Kulkarni1982, Nakanishi2003}. 

By studying the \ion{H}{i} emission at low spatial resolution, \citet{Oort1958} constructed the first face-on \ion{H}{i} distribution map of the Milky Way, and found multiple spiral arms. Later surveys have revealed additional spiral arms in the outer Galaxy \citep[e.g.,][]{Weaver1970, Kulkarni1982, Nakanishi2003, Levine2006}. While most of these works are concentrated on the outer Galaxy, by combining \ion{H}{i} and H$_2$ map derived from CO observations at low angular resolution, \citet{Nakanishi2016} were able to trace the spiral arms from the inner to the outer Galaxy. 

Assuming the 21~cm line is optically thin, the total \ion{H}{i} mass of the Milky Way has been estimated to be 7.2 to 8.0$\times10^9~M_\odot$ \citep{Kalberla2009, Nakanishi2016}. Studies towards nearby gas at high latitudes show that optical depth can be negligible for \ion{H}{i} mass estimation in such a context \citep[e.g.,][]{Lee2015, Murray2018a}, while on the other hand, studies of \ion{H}{i} self absorption towards nearby galaxies (M31, M33 and LMC) revealed that \ion{H}{i} mass can increase by 30--34\% with optical depth correction \citep{Braun2009, Braun2012}. A study toward mini starburst region W43 in our Milky Way revealed an even more extreme correction factor of 240\% for \ion{H}{i} mass estimation when applying optical depth correction. However, no systematic study has yet been done in the Galactic plane.

THOR provides the opportunity to study the distribution and spiral structures of the atomic gas in the northern Galactic plane with the \ion{H}{i} emission data, to further investigate the optical depth, and calibrate the mass estimation of the atomic gas using absorption data.
Furthermore, the atomic hydrogen gas, especially the cold neutral medium (CNM, $T\sim40-100$~K, \citealt{McKee1977, Wolfire1995}) also traces the \ion{H}{i}-to-H$_{2}$ transition. Combining the \ion{H}{i} absorption lines with the simultaneously observed OH absorption lines from the THOR survey and complementary molecular gas information from such as CO observations allows us to study the transition phase between atomic gas and molecular gas \citep{rugel2018}. 

This paper presents the second data release of the THOR survey. We focus scientifically on the \ion{H}{i} emission and absorption. The observation strategy and data reduction details are described in Sect.~\ref{sect_obs}. The parameters of the data products, along with an overview of the \ion{H}{i} results are presented in Sect.~\ref{sect_results}. The \ion{H}{i} results are discussed in Sect.~\ref{sect_discuss}, and our conclusions are summarized in Sect.~\ref{sect_con}. 

%--------------------------------------------------------------------

\section{Observations and Data reduction}
\label{sect_obs}
We observed the first quadrant of the Galactic plane, covering $l=14.0-67.4^\circ$ and $\lvert b \rvert \leq 1.25\degr$ with the VLA in C-configuration in L band from 1 to 2~GHz. The observations were carried out in three phases, a pilot study ($l=29.2\degr-31.5\degr$), phase 1 ($l=14.0\degr-29.2\degr,\ 31.5\degr-37.9\degr$ and $47.1\degr-51.2\degr$), and phase 2 ($l=37.0\degr-47.9\degr$ and $51.1\degr-67.4\degr$), spanning across several semesters (from 2012 to 2014). The detailed observing strategy and data reduction are discussed and described in \citet{Bihr2016}, and data release 1 in \citet{beuther2016}, which present only the pilot and phase I data. With the WIDAR correlator, we cover the \ion{H}{i} 21~cm line, four OH lines ($\Lambda$ doubling transitions of the OH ground state, the ${\rm {}^{2}\Pi_{3/2};J=3/2}$ state, ``main lines'' at 1665 and 1667~MHz, ``satellite lines'' at 1612 and 1720~MHz), 19 H$\alpha$ recombination lines, as well as eight continuum bands, for instance, SPWs. Each continuum SPW has a bandwidth of 128~MHz. Due to strong radio frequency interference (RFI) contaminations, two SPWs around 1.2 and 1.6~GHz were not usable and were discarded. The remaining six SPWs are centered at 1.06, 1.31, 1.44, 1.69, 1.82, and 1.95~GHz. For the fields at $l=23.1-24.3^\circ$ and $25.6-26.8^\circ$, the SPW around 1.95~GHz is also severely affected by RFI and is therefore flagged \citep[see][]{Bihr2016}. Each pointing was observed three times to ensure a uniform $uv$-coverage, and the total integration time is $5-6$~min per pointing. A detailed description of the observational setup can be found in \citet{beuther2016}.

The full survey was calibrated and imaged with the Common Astronomy Software Applications (CASA)\footnote{\url{http://casa.nrao.edu}; version 4.1.0 for the pilot study and phase 1, version 4.1.2 for phase 2.} software package \citep{McMullin2007}. The modified VLA scripted pipeline\footnote{\url{https://science.nrao.edu/facilities/vla/data-processing/pipeline/scripted-pipeline}} (version 1.2.0 for the pilot study and phase 1, version 1.3.1 for phase 2) was used for the calibration. The absolute flux and bandpass were calibrated with the quasar 3C~286. J1822-0938 (for observing blocks with $l<39.1\degr$) and J1925+2106 (for the remaining fields) were used for the phase and gain calibration. Except for the RRLs, all data were inverted and cleaned with multiscale CLEAN in CASA to better recover the large scale structure \citep[see also][]{beuther2016}. 

In most regions, the individual RRLs are too weak to be detected, and so cleaning the image is typically not useful. We therefore stacked the dirty images of all RRL spectral windows that are not affected by RFI with equal weights in the velocity. All RRL dirty images were produced with the same spectral resolution of 10~km~s$^{-1}$ and smoothed to a common angular resolution of 40\arcsec\ before the stacking \citep[see also ][]{beuther2016}.
 
For the continuum and RRL data, we employed the RFlag algorithm in CASA, which was first introduced to AIPS by E. Greisen in 2011 to minimize the effects from RFI in each visibility dataset before imaging. Some SPWs, such as the continuum SPW at $1.2$~GHz and $1.6$~GHz, and some RRLs, have so much RFI over the whole band that no usable data could be recovered by RFlag, and were therefore abandoned \citep[see also][]{beuther2016, Wang2018}.

\subsection{New \ion{H}{i} dataset}
\begin{figure*}
\centering
\includegraphics[angle=0, width=0.8\textwidth]{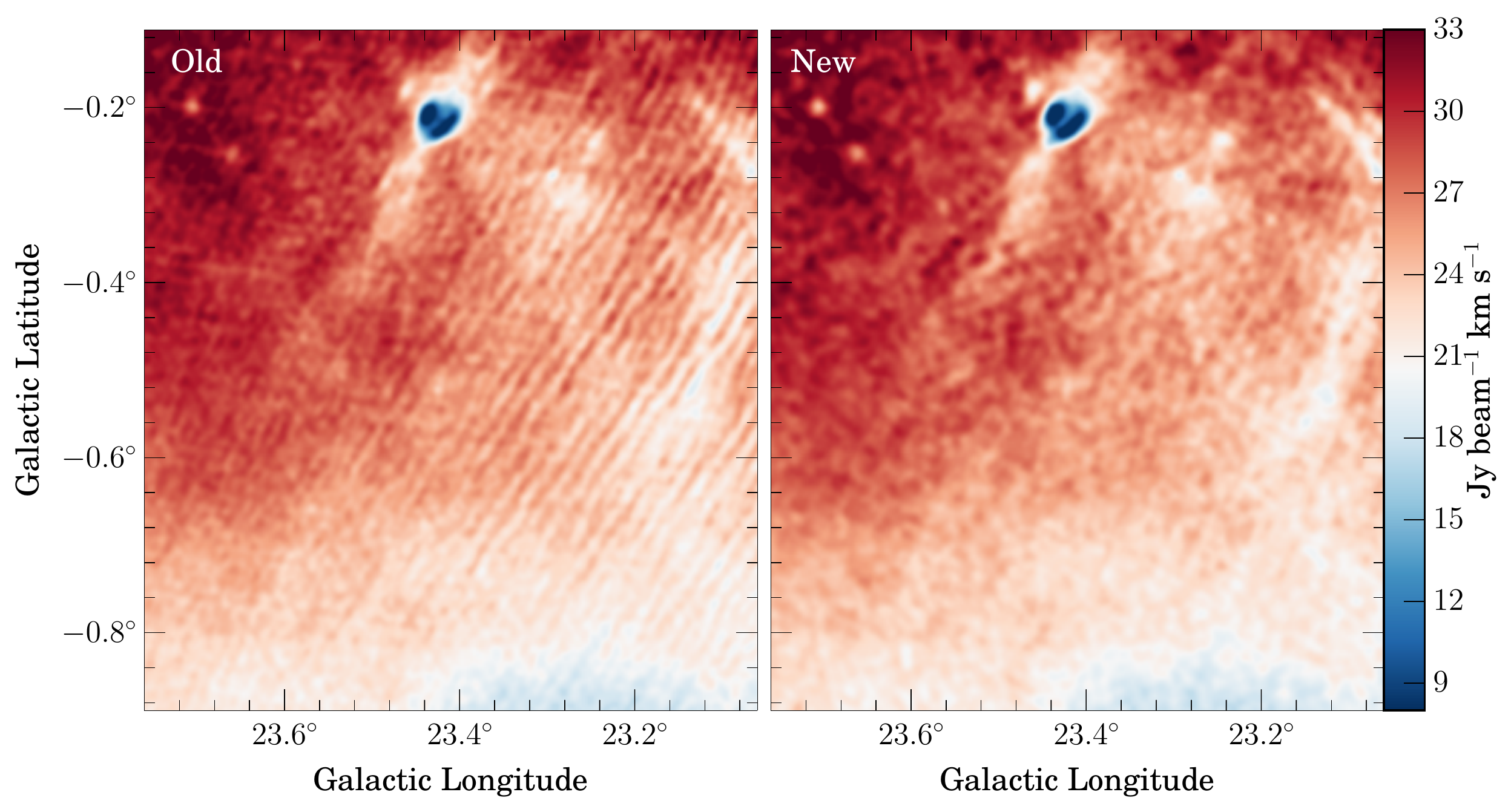} 
\caption{\ion{H}{i} integrated intensity maps (in the velocity range $-100<v_{\rm LSR}<150$~km~s$^{-1}$) for a sample region. The left panel shows the product in the THOR data release 1, which were obtained by direct feathering of the VLA C- and D-configuration observations with the GBT. The right panel shows the product in the THOR data release 2, which were obtained by combining the VLA C- and D-configuration data in the visibility domain and then feathering with the VGPS.}
\label{fig_hi_compare}
\end{figure*}

For the \ion{H}{i} 21~cm line observations, we combined the THOR C-configuration data with the \ion{H}{i} Very Large Array Galactic Plane Survey (VGPS, \citealt{stil2006}), which consists of VLA D-configuration data combined with single-dish observations from the Green Bank Telescope (GBT), to recover the large-scale structure. The \ion{H}{i} data from the pilot and phase 1 of the survey were published in data release 1 \citep{beuther2016}, in which we combined the THOR C-configuration images directly with the published VGPS data using the task ``feather'' in CASA. This method does recover the large scale structure, but the quality of the images is not ideal. The images are quite pixelized and contain many sidelobe artifacts (see left panel in Fig.~\ref{fig_hi_compare}). 

To improve the image quality, we chose a different method to combine the dataset. We first combined the C-configuration data in THOR with the D-configuration of VGPS in the visibility domain. We subtracted the continuum in the visibility datasets with UVCONTSUB in CASA, and used the multiscale CLEAN in CASA\footnote{version 5.1.1} to image the continuum-subtracted C-configuration data together with D-configuration data. The images were afterward combined with the VGPS images (D+GBT) using the task, "feather", in CASA (see also Wang et al. submitted). Since the D-configuration observations of VGPS cover only $l=17.6\degr-67\degr$, the combined \ion{H}{i} data are restricted to $l=17.6\degr-67\degr$, which is slightly smaller than the sky coverage of the C-configuration-only \ion{H}{i} data ($l=14.0-67.4\degr$). Compared to the \ion{H}{i} images from data release 1, the quality of the new images has significantly improved (Fig.~\ref{fig_hi_compare}). More details about the data products are given in Table~\ref{table_product}.

\section{Results}
\label{sect_results}
We describe the parameters of the data products, and present an overview of the \ion{H}{i} results in this section.

\subsection{Data release 2}
All spectral line data from the pilot and phase 1, as well as all Stokes I continuum data, have been published and are already available to the community \citep{bihr2015, Bihr2016, beuther2016, Beuther2019, Walsh2016, rugel2018, Wang2018}. In this paper, we publish the data from the second half of the survey, including the new \ion{H}{I} dataset for the whole survey, available at our project website\footnote{\url{http://www.mpia.de/thor}} and at the CDS\footnote{\url{https://cds.u-strasbg.fr}}. We summarize the basic parameters of the data products in Table~\ref{table_product}. Because of different requirements for calibrating and imaging the polarization data \citep[see also][]{beuther2016}, the data reduction of these data for the whole survey is still ongoing. The first results of the Faraday rotation study in Galactic longitude range 39\degr\ to 52\degr\ are presented by \citet{Shanahan2019}. More polarization data should be available at a later stage.

The noise of our data is dominated by the residual side lobes. Particularly in regions close to strong emission from the continuum and the masers, the noise can increase significantly. The noise properties of the continuum data and OH masers are studied in detail by \citet{Bihr2016}, \citet{beuther2016}, \citet{Walsh2016}, \citet{Wang2018}, and \citet{Beuther2019}. We list only the typical noise values in Table~\ref{table_product}.

\begin{table*}
\caption{Properties of the data products in the THOR data release 2.}
\centering
\label{table_product}
\begin{tabular}{l c c c c c r } 
\hline\hline
    & Rest Freq. & Width            &$\Delta v$          &beam      &  beam        & noise\tablefootmark{a}\\
    &  (MHz)      & (km~s$^{-1}$) & (km~s$^{-1}$)&  native  &  smoothed  & mJy~beam$^{-1}$\\
\hline\vspace{-0.35cm}\\
\ion{H}{I} & 1420.406 & 277.5& 1.5& --& 40\arcsec & 10\tablefootmark{b}\\
\ion{H}{I}+cont. &1420.406 & 300  & 1.5 &13.0\arcsec~to 19.1\arcsec &  25\arcsec & 10\tablefootmark{b}\\ 
OH1 & 1612.231&195& 1.5&11.6\arcsec~to 18.7\arcsec& 20\arcsec & 10\tablefootmark{b}\\ 
OH2 & 1665.402&  195 & 1.5&11.1\arcsec~to 18.1\arcsec& 20\arcsec & 10\tablefootmark{b}\\ 
OH3\tablefootmark{c} & 1667.359&  195 & 1.5&11.0\arcsec~to 13.7\arcsec &20\arcsec & 10\tablefootmark{b}\\ 
OH4 & 1720.530& 195 & 1.5& 11.0\arcsec~to 17.6\arcsec &20\arcsec & 10\tablefootmark{b}\\
RRL\tablefootmark{d} & -- &210&10& -- &40\arcsec & 3.0\tablefootmark{e}\\
cont1\tablefootmark{e} & 1060 &--&--&14.7\arcsec~to 24.4\arcsec&25\arcsec& 1.0\\
cont2\tablefootmark{e} & 1310 &--&--&12.2\arcsec~to 19.7\arcsec&25\arcsec& 0.3\\
cont3\tablefootmark{e} & 1440 &--&--&11.6\arcsec~to 18.1\arcsec&25\arcsec& 0.3\\
cont4\tablefootmark{e} & 1690 &--&--&9.5\arcsec~to 15.4\arcsec&25\arcsec& 0.3\\
cont5\tablefootmark{e} & 1820 &--&--&9.1\arcsec~to 14.5\arcsec&25\arcsec& 0.3\\
cont6\tablefootmark{e} & 1950 &--&--&8.2\arcsec~to 13.1\arcsec&25\arcsec& 0.7\\
cont3+VGPS &1420& --&--&--&25\arcsec & 6.5\\
\hline   
\end{tabular} 
\tablefoot{
\tablefoottext{a}{Typical noise of the data with the smoothed beam. Due to residual side lobes, the noise can be higher in areas round strong sources.}
\tablefoottext{b}{Noise of the line data is measured in the line free channels.}
\tablefoottext{c}{The OH line at 1667~MHz is observed for $l=29.2\degr-31.5\degr,\ 37.9\degr-47.9\degr,$ and $51.1\degr-67.0\degr$.}
\tablefoottext{d}{RRL images were produced by smoothing and stacking all available RRL images.}
\tablefoottext{e}{Each continuum band has a bandwidth of 128~MHz.}
}
\end{table*}

\subsection{OH}
\label{oh}
Continuum subtracted images at both the native and the smoothed resolution (20\arcsec) of the four OH lines are provided to the community. The correlator setup was slightly different between the pilot study, phase 1, and phase 2, which mainly affects the sky coverage of OH lines and RRLs and the native spectral resolution of spectral lines (see \citealt{beuther2016} for details). The OH line at 1667~MHz was only observed in the pilot study and phase 2 ($l=29.2\degr-31.5\degr,\ 37.9\degr-47.9\degr$, and $51.1\degr-67.0\degr$, see also, \citealt{rugel2018} and \citealt{Beuther2019}). 

Diverse physical processes are traced by OH masers, from expanding shells around evolved star to shocks produced by star-forming jets or SNRs  \citep{Elitzur1992}. \citet{Beuther2019} survey identified 1585 individual maser spots distributed over 807 maser sites in the THOR survey, among which $\sim$50\% of the maser sites are associated with evolved stars, $\sim20$\% are associated with star-forming regions, and $\sim3$\% are potentially associated with SNRs (see \citealt{Beuther2019} for details).

Thermal OH lines are often detected as absorptions towards strong continuum sources, and they can be used to trace molecular clouds where CO is not detected, so called ``CO-dark'' regions \citep[e.g.,][]{Allen2015, Xu2016}. By studying the two main transitions (1665 and 1667~MHz), \citet{rugel2018} detect 59 distinct OH absorption features against 42 continuum background sources, and most of the absorptions occur in molecular clouds associated with Galactic \ion{H}{ii} regions. This is the first unbiased interferometric OH survey towards a significant fraction of the inner Milky Way, and provides a basis for theoretical and future follow-up studies (see \citealt{rugel2018} for details).
 
\subsection{RRLs}
\label{rrl}
As mentioned in the previous section, to increase the signal-to-noise (S/N), ratio we smoothed and stacked all available recombination line images. The final RRL images have an angular resolution of 40\arcsec\ and a velocity resolution of 10~km~s$^{-1}$. The typical linewidths of RRLs measured toward \ion{H}{ii} regions are around $\sim20-25$~km~s$^ {-1}$\citep{Anderson2011}, so a 10~km~s$^{-1}$ resolution is reasonable to study the kinematics of \ion{H}{ii} regions.
 
Combining the RRL data from THOR with complementary CO data, \citet{Rugel2019} found shell-like structures in RLL emission toward W49A. The ionized emission and molecular gas emission show correlation towards the shell-like structures. By comparing to one-dimensional feedback models \citep{Rahner2017, Rahner2019}, \citet{Rugel2019} suggest W49A is potentially regulated by feedback-driven and re-collapsing shells (see \citealt{Rugel2019} for details).

Mostly due to sensitivity limitations, interferometric mapping surveys of RRL emission have been rare \citep[e.g.,][]{Urquhart2004}. The stacking method allows us to achieve higher sensitivities than is usually possible when only single lines are observed. THOR provides the community with a new set of RRL maps towards a large sample of \ion{H}{ii} regions, which can be used for sample statistical studies.

\subsection{Continuum}
\label{continuum}
As presented in \citet{Wang2018}, the THOR survey provides the continuum data (both at the native resolution and smoothed to a resolution of 25\arcsec), as well as a continuum source catalog, to the community. To recover the extended structure, we combined the C-configuration 1.4~GHz continuum data from THOR with the 1.4~GHz continuum data from the VGPS survey (D+Effelsberg) using the task, ``feather'' \citep{Wang2018}. The resulting images for the pilot region and phase 1 are very similar to the ones obtained using the VGPS continuum as an input model in the deconvolution of the THOR data. This combined dataset retains the high angular resolution of the THOR observations (25\arcsec), and at the same time, it can recover the large-scale structure. \citet{anderson2017} identified 76 new Galactic SNR candidates in the survey area with this dataset. \citet{anderson2017} further showed that despite the different bandwidths between the VGPS continuum ($\sim1$~MHz) and the THOR continuum ($\sim128$~MHz), the flux retrieved from the combined data is consistent with the literature.

The continuum source catalog contains 10~387 objects that we extracted across our survey area. With the extracted peak intensities of the six usable SPWs between 1 and 2 GHz, we were able to determine a reliable spectral index (spectral index fitted with at least 4 SPWs) for 5657 objects. By cross-matching with different catalogs, we found radio counter parts for 840 \ion{H}{ii} regions, 52 SNRs, 164 planetary nebulae, and 38 pulsars. A large percentage of the remaining sources in the catalog are likely to be extragalactic background sources, based on their spatial and spectral index distributions. A detailed presentation of the continuum catalog can be found in \citet{Bihr2016} and \citet{Wang2018}.

\subsection{\ion{H}{i}}
For the \ion{H}{i} 21~cm line, in addition to the data cubes from C+D+GBT data, we also provide the C-configuration-only \ion{H}{i} datacubes with continuum at both the native resolution and the smoothed beam (25\arcsec), which can be used to measure the \ion{H}{i} optical depth towards bright background continuum sources \citep{Bihr2016}.

A case study of \ion{H}{i} self-absorption towards a giant molecular filament is presented by Wang et al. (submitted). In the following sections, we focus scientifically on \ion{H}{i} emission and absorption.

\subsubsection{\ion{H}{i} emission}
\label{sect_hi}

\begin{figure*}[htbp]
\centering
\includegraphics[angle=90, height=23.8cm]{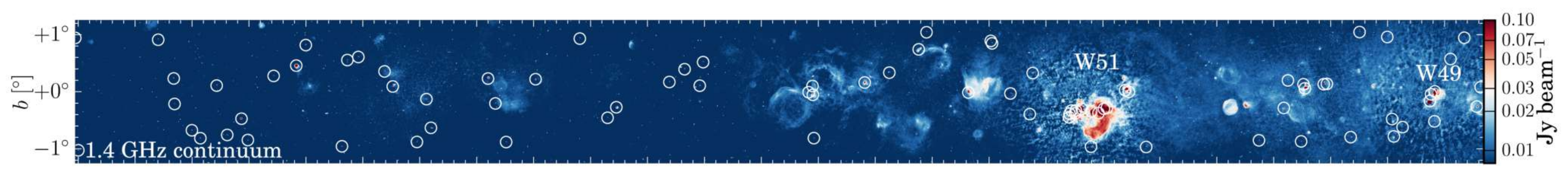} 
\includegraphics[angle=90, height=24cm]{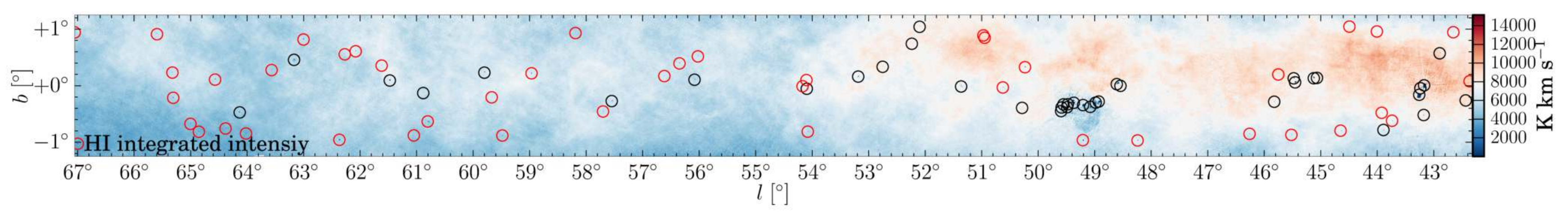} 
\includegraphics[angle=90, height=23.8cm]{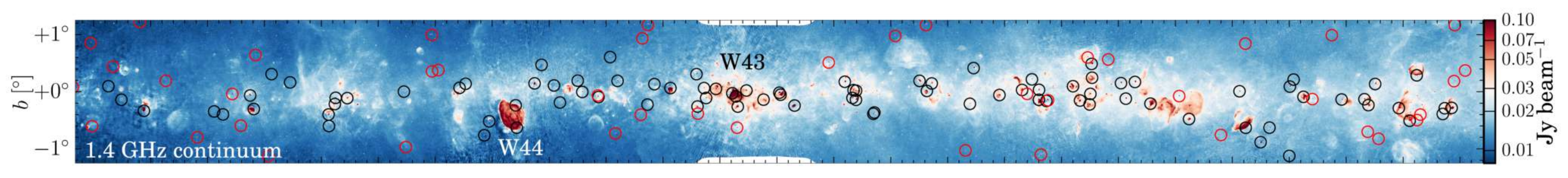} 
\includegraphics[angle=90, height=24cm]{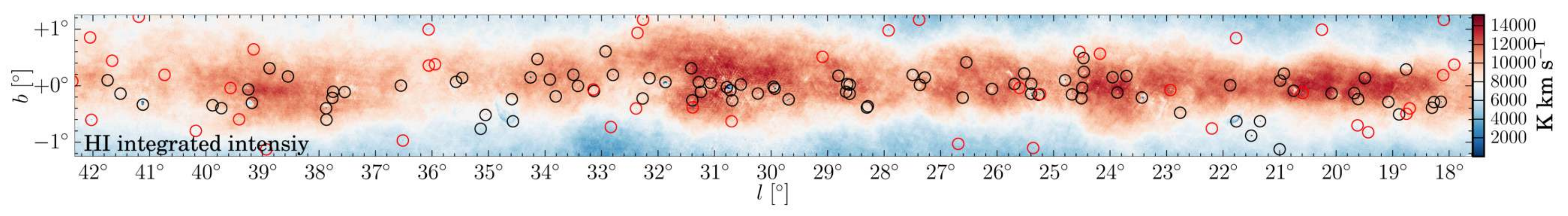} 

\caption{THOR+VGPS 1.4~GHz continuum map and \ion{H}{i} integrated intensity map ($-113<v_{\rm{LSR}} <163$\,km\,s$^{-1}$). The continuum map has a beam size of 25\arcsec, and the beam size for the \ion{H}{i} map is 40\arcsec. In all panels, the circles mark the continuum sources that we used to extract the \ion{H}{i} absorption spectra and construct the optical depth map (see Sect.~\ref{sec_tau}). The Galactic sources are marked with white and black circles, and the extragalactic sources are marked with red circles. A few well-known Galactic sources are also labeled in each panel (W43 at $l\sim31$\degr, W44 $l\sim35$\degr, W49 $l\sim43$\degr and W51 $l\sim49$\degr).}
\label{fig_himom0}
\end{figure*}

Figure~\ref{fig_himom0} shows the C+D+single-dish 1.4~GHz continuum map and the \ion{H}{i} integrated intensity map ($v_{\rm LSR}=-113\sim163$~km s$^{-1}$). While the combined \ion{H}{i} emission data illustrates that the atomic gas is confined near the Galactic mid-plane at lower longitudes ($l<42\degr$), at larger longitudes ($l>54\degr$) gas is more evenly distributed in latitudes. The \ion{H}{i} map in Fig.~\ref{fig_himom0} also shows absorption patterns against some strong continuum sources, such as the star-forming complex W43 at $l\sim30.75\degr$, and W49 at $l\sim43\degr$.

\begin{figure*}[htbp]
\centering
\includegraphics[angle=90,height=24cm]{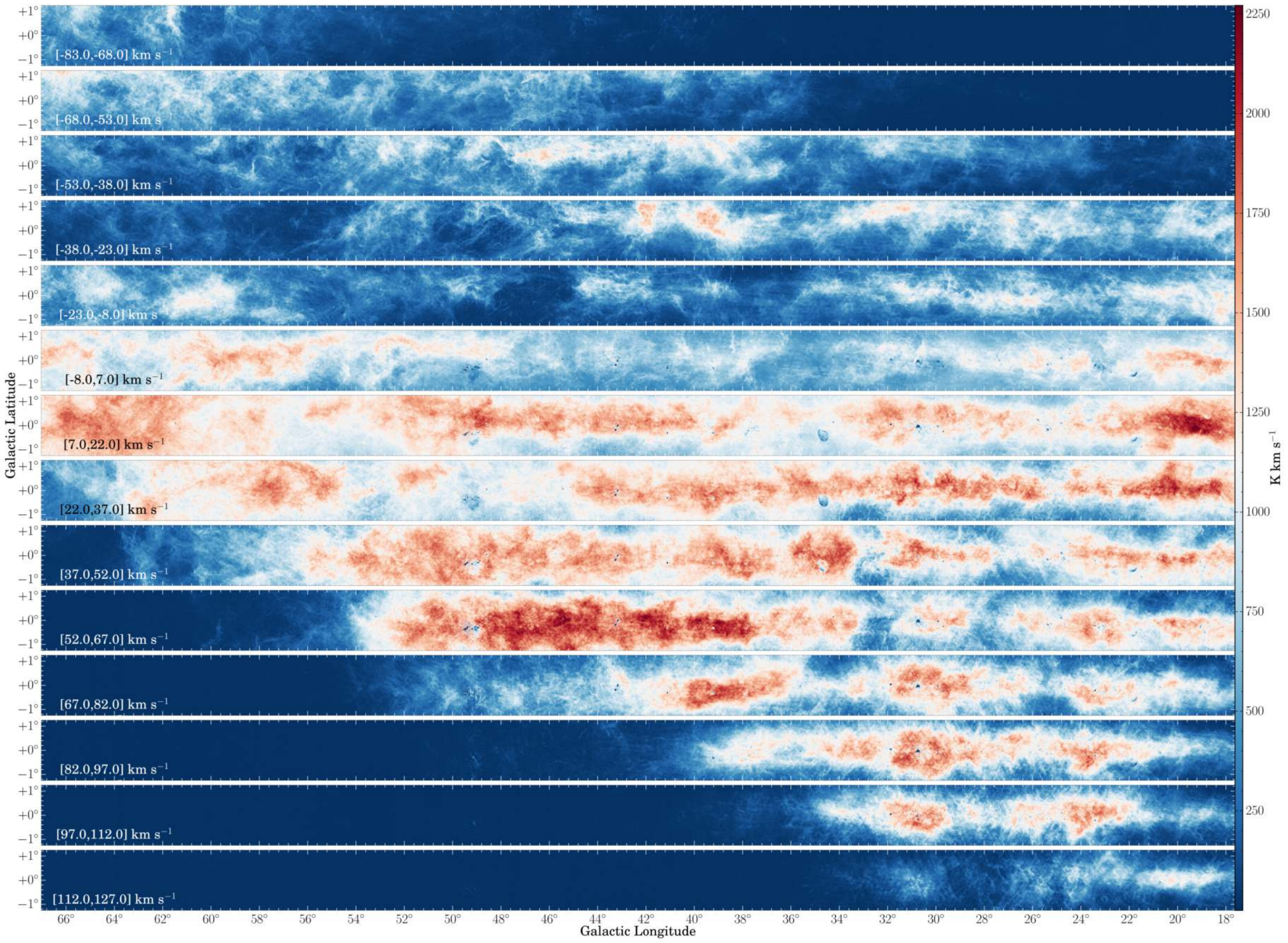} 
\caption{\ion{H}{i} channel maps produced by integrating the \ion{H}{i} emission over indicated velocity ranges. Strong absorption towards W43 ($l\sim31$\degr), W44 ($l\sim35$\degr), W49 ($l\sim43$\degr) and W51 ($l\sim49$\degr) are clearly visible across several velocity channels.}
\label{fig_hi_channel}
\end{figure*}

Figure~\ref{fig_hi_channel} depicts the channel maps by integrating the \ion{H}{i} emission over 15~km~s$^{-1}$ velocity bins. The neutral gas is clearly seen located at higher latitude in channels at $v_{\rm LSR}<-38$~km~s$^{-1}$. This is likely due to the \ion{H}{i} disk being strongly warped in the outer Milky Way \citep{Burton1986, Diplas1991, Nakanishi2003, Kalberla2009}. Some cloud structures are more prominent in the channel maps, such as the filamentary structures in the channels at $v_{\rm LSR}<-38$~km~s$^{-1}$. While at negative velocities the filamentary structures can been seen clearly, the emission at positive velocities appears less structured. This observational phenomenon is likely due to higher spatial and velocity crowding of structures at positive velocities. The negative velocity channels are tracing the outer Galactic plane, and there is little line of sight confusion, while at positive velocities, due to near-far distance ambiguity, the emission from multiple spiral arms could be at the same velocity and result in structures at different distances bing blended together. 

The integrated intensity map (Fig.~\ref{fig_himom0}) also demonstrates that the angular size of the \ion{H}{i} emission along the latitude is smaller at lower longitudes than at larger longitudes. This is due to the fact that the tangent points are further away at lower longitudes. This is also seen in the channel map (Fig.~\ref{fig_hi_channel}) in the positive velocities, in which the tangent points are traced by the left end of the emission in each panel with positive velocities, and the emission structures appear to be smaller at lower longitudes due to them being further away \citep[see also,][]{Merrifield1992}.

\begin{figure*}
\centering
\includegraphics[width=\textwidth]{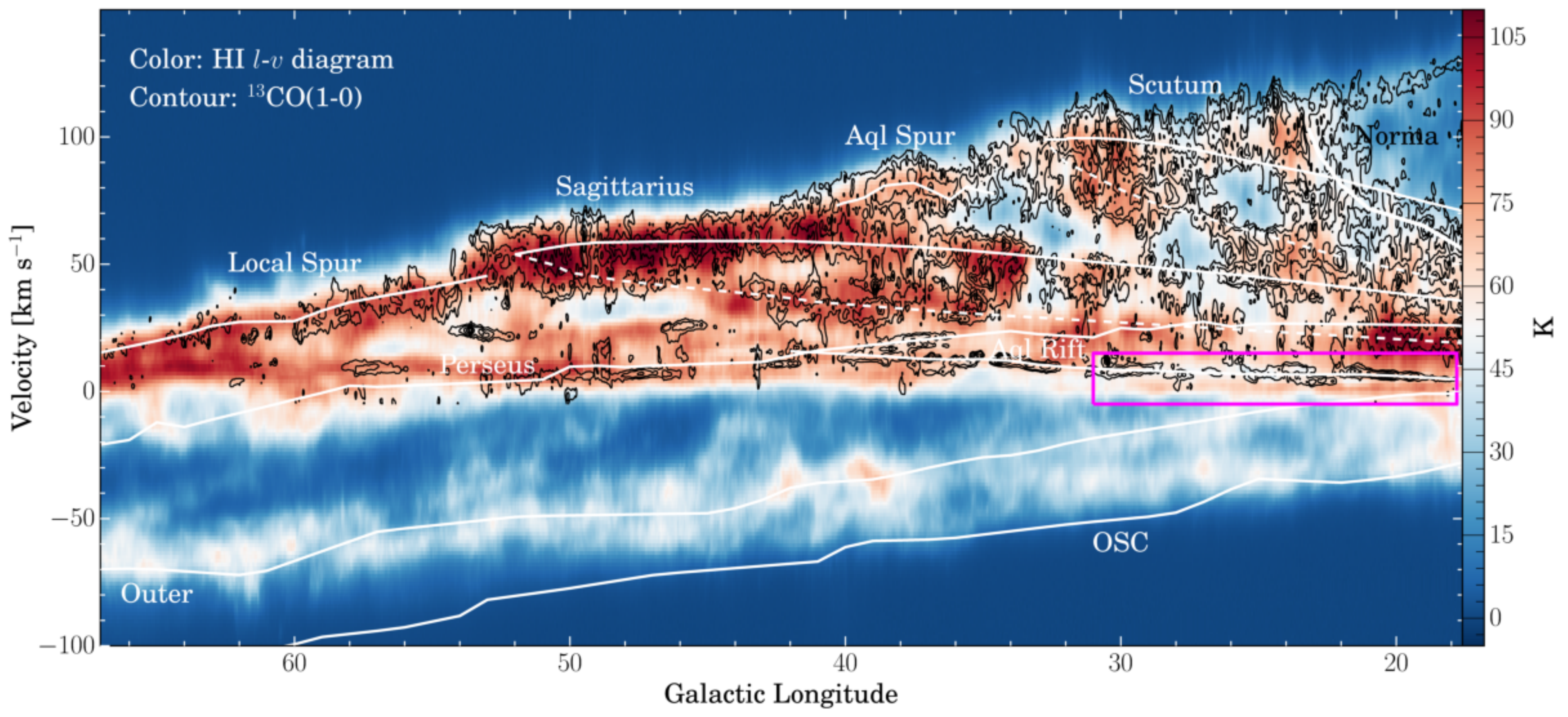}
\caption{\ion{H}{i} longitude-velocity ($l-v$) diagram constructed by averaging the emission in the latitude range $\lvert b \rvert \leq 1.25\degr$. The black contours correspond to the $^{13}$CO(1--0) $l-v$ at 3, 8, 18, 28, and 38 times the rms noise of the $l-v$ diagram (rms = 0.1~K, $T_{\rm MB}$). The $^{13}$CO(1--0) data do not cover the velocity range $v_{\rm LSR}<-5$~km~s$^{-1}$. The overlaid curves trace the Sagittarius, Scutum, Norma, Perseus, Outer, and Outer Scutum-Centaurus (OSC) arms, and smaller features (Local Spur, Aquila Spur, and Aquila Rift) taken from \citet{Reid2016}. The near and far sides of the arms are plotted with dashed and solid lines, respectively. The magenta box marks the absorption feature could be caused by the Riegel-Crutcher cloud. The $^{13}$CO(1--0) data are from the Exeter FCRAO CO Galactic Plane Survey \citep{Mottram2010}, and the GRS \citep{Jackson2006}.}
\label{fig_hi_pv}
\end{figure*}

By averaging the emission in the latitude axis, we constructed the longitude-velocity ($l-v$) diagram of the \ion{H}{i} emission (Fig.~\ref{fig_hi_pv}). The general shape of the $l-v$ diagram agrees with one obtained with the VGPS data \citep{Strasser2007}, but the higher resolutions reveal much finer details, such as the vertical lanes at $l\sim31\degr,\ 43\degr$, and $\sim49\degr$, which are caused by absorption against the background continuum emission from the star-forming regions W43, W49, and W51, respectively (see also Fig.~\ref{fig_himom0}). We also created the $^{13}$CO(1--0) $l-v$ diagram with the same method and plotted it as contours in Fig.~\ref{fig_hi_pv}. The $^{13}$CO(1--0) data is taken from the Exeter FCRAO CO Galactic Plane Survey \citep{Mottram2010}, and Galactic Ring Survey \citep[GRS,][]{Jackson2006}. We re-gridded the Exeter $^{13}$CO data to the same velocity resolution and coverage as the GRS, so there is no $^{13}$CO coverage at velocities where $v_{\rm LSR}<-5$~km~s$^{-1}$. Compared to the $^{13}$CO emission, the \ion{H}{i} emission also traces more extended and diffuse structures.

The \ion{H}{i} emission in general agrees well with the spiral arm models of \citet{Reid2016}, except for the Outer Scutum-Centaurus (OSC) Arm. The OSC Arm is at a large Galactocentric distance and is outside of our survey area at longitudes larger than 40\degr\ due to the warping and flaring of the outer disk \citep{Dame2011, Armentrout2017}. Thus, we can only detect a small portion of the OSC Arm in the inner part of the Galactic plane that our survey covers, as illustrated in Fig.~\ref{fig_hi_pv}.
 
Another feature in the $l-v$ diagram is a strong self-absorption pattern at $\sim$5~km~s$^{-1}$ following the Aquila Rift (magenta box in Fig.~\ref{fig_hi_pv}). This absorption feature spans almost 20\degr\ in longitude ($\sim$17\degr\ to 36\degr). The $^{13}$CO emission, on the other hand, lies right inside the absorption feature in the $l-v$ diagram. This large-scale absorption feature could be caused by the Riegel-Crutcher cloud, which centers at $v_{\rm LSR}\sim5$~km~s$^{-1}$ and covers the longitude range $l=345\degr$ to 25\degr,\ and the latitude range $b\leq6\degr$ \citep{Riegel1969, Riegel1972, Crutcher1974}.

\begin{figure*}
\centering
\includegraphics[width=\textwidth]{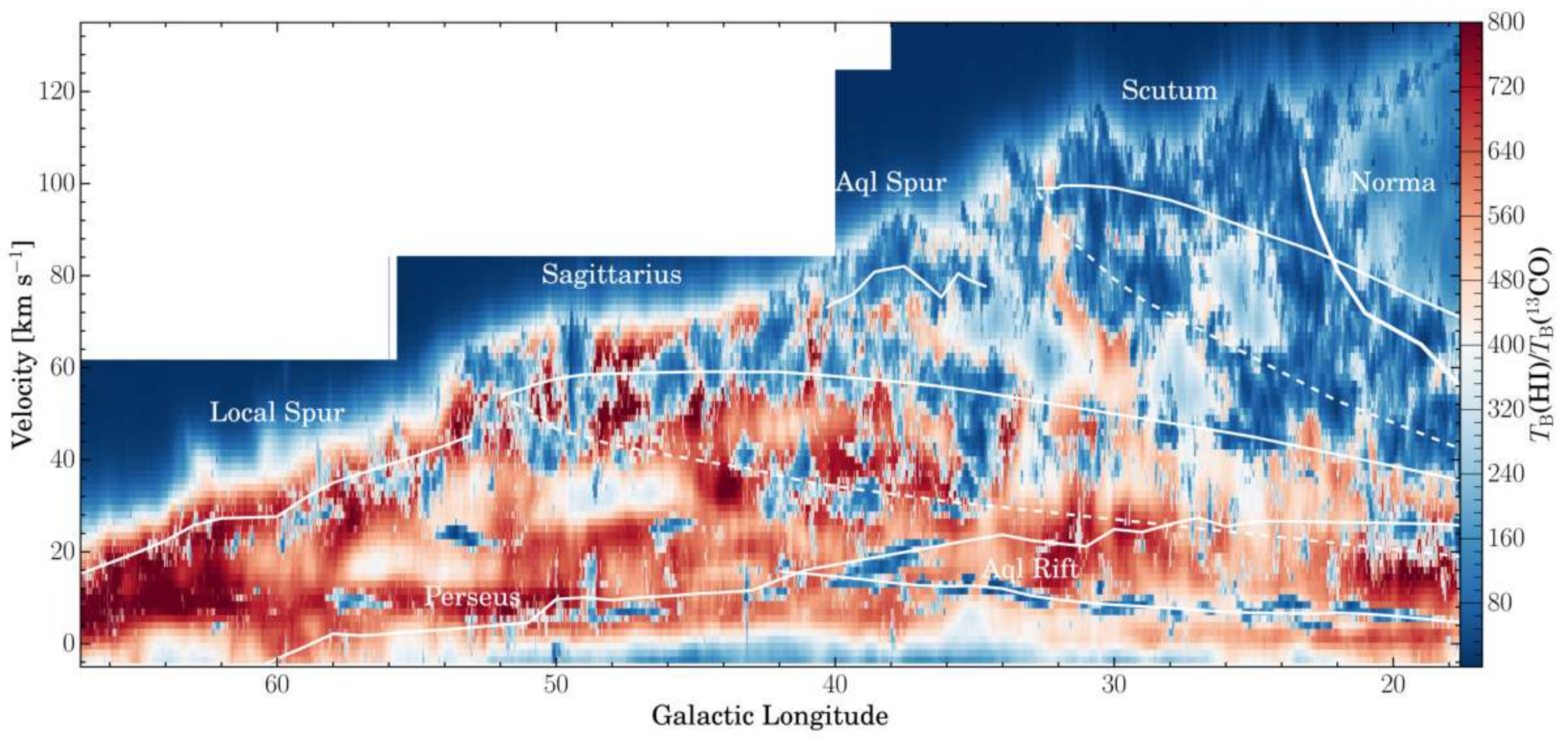}
\caption{The ratio map $l-v$ diagram of \ion{H}{i} over $^{13}$CO ($T_{\rm B}$(\ion{H}{i})/$T_{\rm B}$($^{13}$CO)) averaged over galactic latitude. The $^{13}$CO(1--0) data do not cover the velocity range $v_{\rm LSR}<-5$~km~s$^{-1}$. The overlaid curves trace the Sagittarius, Scutum, Norma, Perseus, Outer, and Outer Scutum-Centaurus (OSC) arms, and smaller features (Local Spur, Aquila Spur, and Aquila Rift) taken from \citet{Reid2016}. The near and far sides of the arms are plotted with dashed and solid lines, respectively.}
\label{fig_ratio_pv}
\end{figure*}

After resampling both the \ion{H}{i} and $^{13}$CO data to the same angular and spectral resolution (pixel size of 22\arcsec, beam size of 46\arcsec, and velocity resolution of 1.5~km~s$^{-1}$), we constructed the $T_{\rm B}$(\ion{H}{i})/$T_{\rm B}$($^{13}$CO) ratio $l-v$ diagram by dividing the \ion{H}{i} $l-v$ diagram by the $^{13}$CO $l-v$ diagram (Fig.~\ref{fig_ratio_pv}). For the $^{13}$CO $l-v$ diagram, a 3$\sigma$ value was used where the emission is below 3$\sigma$ (1$\sigma=0.04$~K in the $^{13}$CO $l-v$ diagram). As expected, Fig.~\ref{fig_ratio_pv} shows that the $T_{\rm B}$(\ion{H}{i})/$T_{\rm B}$($^{13}$CO) ratio is low where there is $^{13}$CO emission (see also, Sect~\ref{sect_hi_ratio}). 

\begin{figure}
\centering
\includegraphics[width=0.5\textwidth]{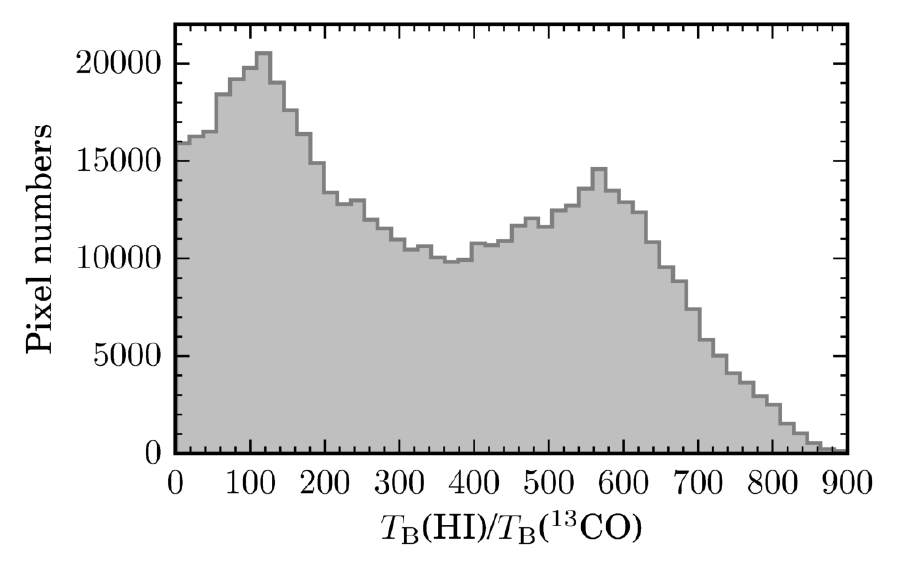}
\caption{Histogram of the $T_{\rm B}$(\ion{H}{i})/$T_{\rm B}$($^{13}$CO) ratio $l-v$ diagram shown in Fig.~\ref{fig_ratio_pv}.}
\label{fig_ratio_hist}
\end{figure}

After removing the pixels with no \ion{H}{i} emission ($T_{\rm B}$(\ion{H}{i})$<5\sigma$, 1$\sigma$=0.2~K), the histogram of the $T_{\rm B}$(\ion{H}{i})/$T_{\rm B}$($^{13}$CO) ratio $l-v$ diagram in Fig.~\ref{fig_ratio_hist} reveals one strong peak at $\sim100,$ and a secondary peak at $\sim600$. If we assume both \ion{H}{i} and $^{13}$CO emissions are optically thin and uniform excitation temperature, the variations in the $T_{\rm B}$(\ion{H}{i})/$T_{\rm B}$($^{13}$CO) ratio also represent the variations in the atomic to molecular gas column density ratio. Figure~\ref{fig_ratio_hist} indicates that the atomic-to-molecular gas column density ratio can increase by a factor of six from inter-arm regions to spiral arms. Considering that we used 3$\sigma$ value of the $^{13}$CO $l-v$ diagram for regions where there is no $^{13}$CO emission to construct the $T_{\rm B}$(\ion{H}{i})/$T_{\rm B}$($^{13}$CO) ratio $l-v$ diagram, this factor of six is a lower limit (see also Sect~\ref{sect_hi_ratio}).

\subsubsection{\ion{H}{i} optical depth}
\label{sec_tau}
\begin{figure*}
\centering
\includegraphics[width=0.3\textwidth]{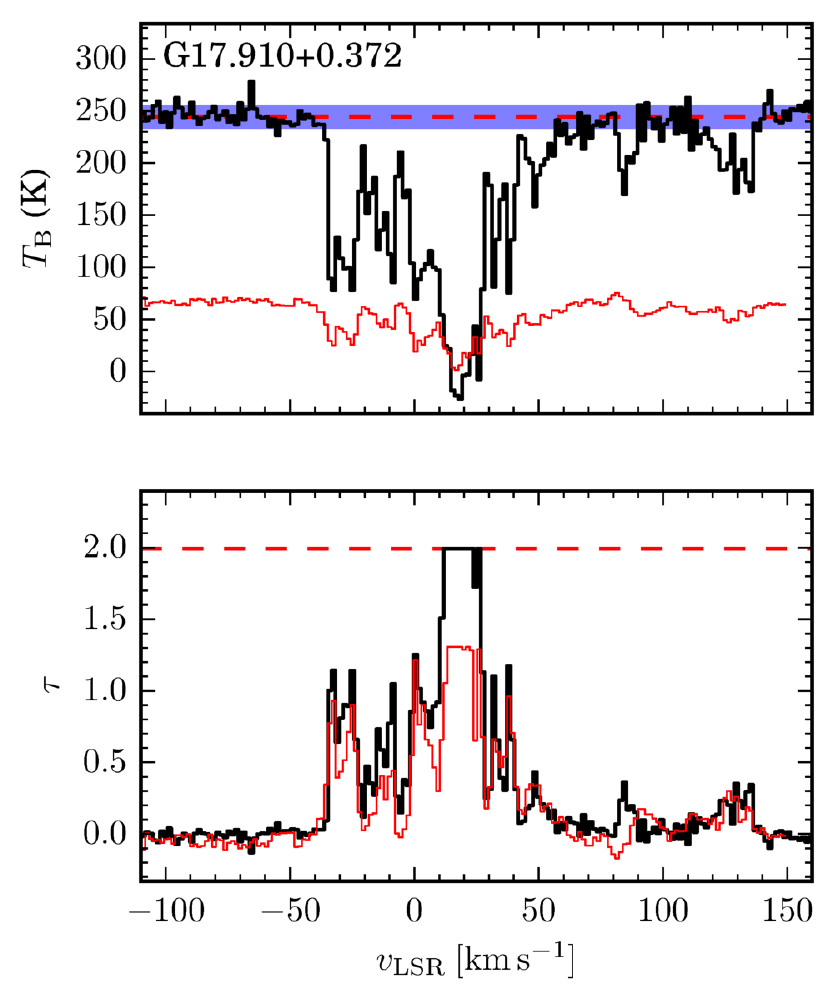}
\includegraphics[width=0.3\textwidth]{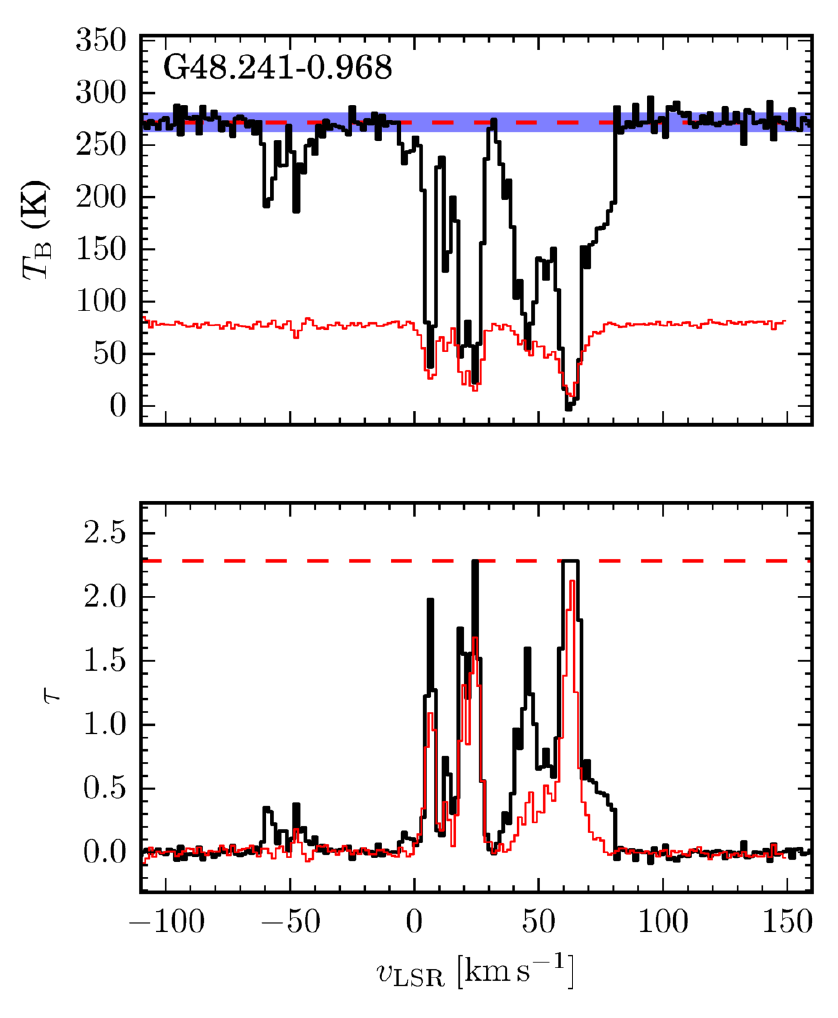}
\includegraphics[width=0.3\textwidth]{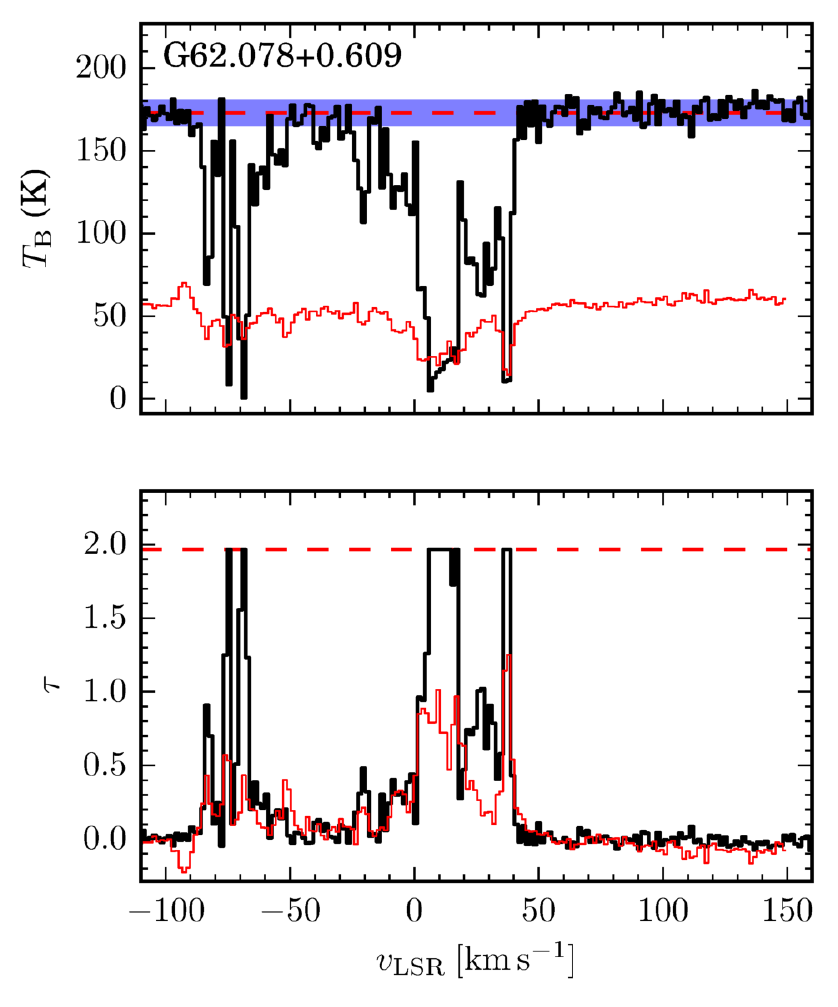}
\caption{\ion{H}{i} absorption and optical depth ($\tau$) spectra of THOR (thick black spectra) and VGPS (thin red spectra). The dashed lines mark the continuum level $T_{\rm cont}$ in the top panels, and the lower limit of $\tau$ in the bottom panels. The shaded bands in the top panels mark the 1$\sigma$ level of the noise.}
 \label{fig_spectra}
\end{figure*}

The THOR C-configuration-only \ion{H}{i} data \citep{beuther2016}, which have not been continuum-subtracted, are used to measure the \ion{H}{i} optical depth towards bright background continuum sources. We extracted the \ion{H}{i} spectra towards all continuum sources with an S/N ratio larger than seven from \citet{Wang2018}. Since the synthesized beam of the C-configuration-only \ion{H}{i} data is 25\arcsec, we extracted the average \ion{H}{i} spectrum from a $28\arcsec\times28\arcsec$ ($7\times$7 pixels) area centered on the position of the continuum source to increase the S/N ratio. We used the software STATCONT \citep{Sanchez-Monge2018} to estimate the continuum level $T_{\rm cont}$ and the noise of the spectra. Some example spectra are shown in Fig.~\ref{fig_spectra}. With the absorption spectra, we can estimate the optical depth towards the continuum source following the method described in \citet{bihr2015}:
\begin{equation}
\tau = -{\rm ln}\left(\frac{T_{\rm{on,\ cont}} - T_{\rm off,\ cont}}{T_{\rm{cont}}}\right),
\label{eq_tau}
\end{equation}
%-
where $T_{\rm on,\ cont}$ is the brightness temperature of the absorption feature measured towards the continuum source, and $T_{\rm off,\ cont}$ is the off-continuum-source brightness temperature. Since we use the THOR C-configuration data to calculate $\tau$, the smooth, large-scale structure is mostly filtered out \citep{beuther2016}. Therefore, we can neglect the off emission $T_{\rm off,\ cont,}$ and simplify Eq.~\ref{eq_tau} to:
%-
\begin{equation}
\tau_{\rm{simplified}} = -{\rm ln}\left(\frac{T_{\rm{on,\ cont}}}{T_{\rm{cont}}}\right) .
\label{eq_tau_simplified}
\end{equation}
%-

For channels with a $T_{\rm{on,\ cont}}$ value less than three times the rms, we use the 3$\sigma$ value to get a lower limit on $\tau$. The bottom panels in Fig.~\ref{fig_spectra} show the $\tau$ spectra for the corresponding sources, and the calculated $\tau$ is always saturated in some channels. Compared to the VGPS VLA D-configuration absorption spectra and the $\tau$ spectra derived from them (resolution 60\arcsec, \citealt{Strasser2007}), the THOR C-configuration absorption spectra are much more sensitive and therefore probing higher optical depth (Fig.~\ref{fig_spectra}).

\begin{figure*}
\centering
\includegraphics[width=\textwidth]{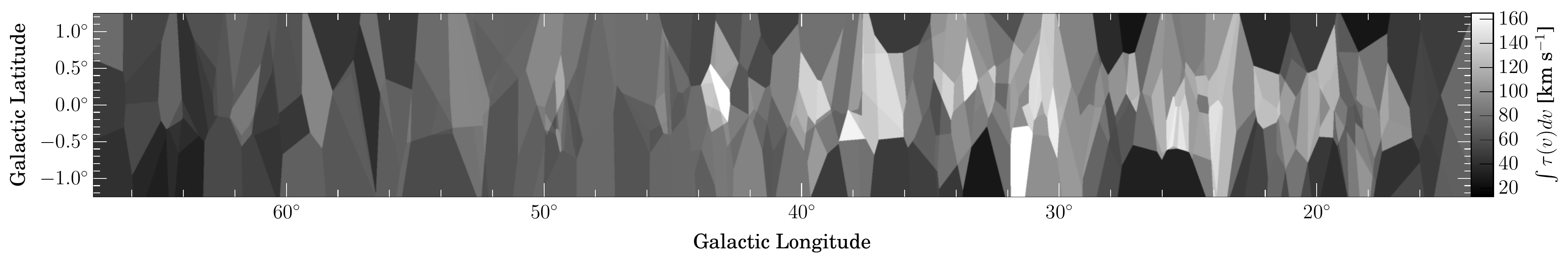}
\caption{\ion{H}{i} integrated optical depth ($\tau$) map across the velocity range ($-128.5<v_{\rm LSR}<$170~km~s$^{-1}$). The aspect ratio of the figure is modified for demonstration purpose.}
 \label{fig_tau_map}
\end{figure*}

In this study, since only $T_{\rm{on,\ cont}}<(T_{\rm cont} - 3~\sigma$) is considered to be real absorption, only sources with $T_{\rm cont}>6~\sigma$ can have channels with real absorption and do not reach the lower limit of $\tau$. In total, 228 sources have a $T_{\rm cont}>6~\sigma$ to make the $\tau$ map, among which $\sim$60\% are Galactic sources (Table~\ref{table_tau}). We listed $T_{\rm cont}$, $\sigma$ of $T_{\rm cont}$, the lower limit of $\tau$, and the integrated $\tau$ including the physical nature of the 228 sources in Table~\ref{table_tau}. We then grid the 228 $\tau$ measurements channel by channel for the whole survey using the nearest-neighbor method\footnote{\url {https://docs.scipy.org/doc/scipy/reference/generated/scipy.interpolate.griddata.html}} to create the $\tau$ data-cube. Since the Galactic sources do not trace any absorption from the gas located behind them, we replaced the optical depth values for these channels with the ones from the nearest extragalactic sources. The resulting integrated $\tau$ map is shown in Fig.~\ref{fig_tau_map}. 

The highest integrated $\tau$ we measured is at $l\sim43$\degr, at the location of the high-mass star-forming complex W49A. A very high integrated $\tau$ is measured towards the star-forming complex W43 ($l\sim31$\degr). Figure~\ref{fig_tau_map} shows that higher $\tau$ is measured in the inner longitude range, which is reasonable considering more material is packed along the line of sight due to velocity crowding in regions below $l\sim43$\degr. Since the used $\tau$ spectra are all saturated in some channels, this map represents a lower limit of the optical depth in the survey region.

\subsubsection{\ion{H}{i} column density and distribution}
\label{sect_hi_dist}
We estimated the column density of atomic hydrogen from the emission line data (C+D+GBT) using \citep[e.g., ][]{wilson2013}:
%-
\begin{equation}
N_{\rm H} = 1.8224\times10^{18}\: \int T_{\rm S}(v) \tau(v)\, dv.
\label{eq_column_density_hi}
\end{equation}
%-

The optical depth corrected spin temperature is $T_{\rm S}(v)=T_{\rm B}(v)/(1-{\rm e}^{-\tau(v)})$, where $T_{\rm B}$ is the brightness temperature of the \ion{H}{i} emission. We used the $\tau$ data-cube (Sect.~\ref{sec_tau}) to correct the spin temperature channel by channel and estimate the column density. 

\begin{figure}
\centering
\includegraphics[width=0.5\textwidth]{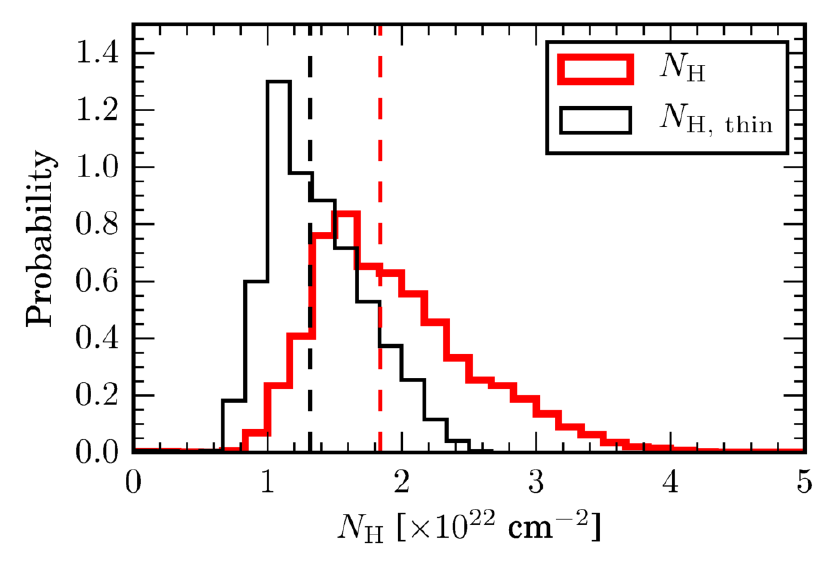}
\caption{Histograms of the atomic hydrogen column density integrated between --113 to 163~km~s$^{-1}$. The black (thin) histogram represents the column density derived assuming the emission is optically thin, and the red (thick) histogram represents the column density corrected fro optical depth. The dashed vertical lines mark the median values of the two histograms, respectively. }
 \label{fig_nh_histo}
\end{figure}

As shown in Fig.~\ref{fig_himom0} and Fig.~\ref{fig_hi_channel}, \ion{H}{i} absorption against strong continuum sources is clearly seen as negative features in the continuum subtracted emission map, such as toward W43 at $l\sim31\degr$. To derive the column density toward strong continuum sources, we determined a mean brightness temperature from a region of radius 10~\arcmin\ around the continuum source to derive $T_{\rm S}$. Then, we extracted C-configuration-only \ion{H}{i}+continuum spectra from these regions pixel-by-pixel to derive the optical depth (see Sect.~\ref{sec_tau}). With the spin temperature and optical depth derived, we can estimate the column density towards continuum sources and add these to the optical depth corrected column density map. Figure~\ref{fig_nh_histo} shows the histograms of the atomic hydrogen column density integrated between --113 to 163~km~s$^{-1}$ of the survey area. The median value of the column density is 1.8$\times10^{22}$~cm$^{-2}$, which is 38\% higher than the value assuming optically thin emission.

To obtain the distribution of atomic gas in the Galactic plane, we estimated the kinematic distance of each channel and each pixel of the \ion{H}{i} emission map with the Kinematic Distance Utilities\footnote{\url{https://github.com/tvwenger/kd}} \citep{Wenger2018}. The ``universal'' rotation curve of \citet{Persic1996} including terms for an exponential disk and a halo, as well as the Galactic parameters from \citet{Reid2014} are used. See Table.~5 in \citet{Reid2014} for detailed parameters. 

To solve the kinematic distance ambiguity in the inner Galaxy (inside the solar orbit), we took the following approach. We assume that the average vertical density profile of the atomic gas $n(z)$ in the inner Galaxy can be described by the sum of two Gaussians and an exponential function \citep{Lockman1984, Dickey1990}:
%-
\begin{equation}
\begin{split}
n(z) =  \ & \sum_{i=1}^2 n_i(0)\ {\rm exp}\ \left[-z^2/\left(2h_i^2\right)\right]\\
             &+ n_3(0)\ {\rm exp}\ \left(-|z|/h_3\right),
\end{split}
\label{eq_hi_vertical}
\end{equation}
%-
with $z$ describing the distance from the Galactic mid-plane. The coefficients from \citet{Dickey1990} are listed in Table~\ref{table_hi_vertical}. Since the volume density distribution of the atomic gas in the mid-plane is approximately axisymmetric with respect to the Galactic center \citep{Kalberla2008}, we can assume the volume density is the same at the same Galactocentric distance in the mid-plane. Furthermore, the $v_{\rm LSR}$ distance profiles are symmetric with respect to the tangent point for the degeneracy part \citep[see also,][]{Anderson2009}, so each velocity bin traces the same line-of-sight distance at the near side as at the far side. Thus we assume the column density is also the same at the same Galactocentric distance at the near and far side in the Galactic mid-plane. Due to the kinematic distance ambiguity in the inner Galaxy, the column density we derived for each pixel at each velocity channel is a combined result from both the far and near side. Thus, we can use Eq.~\ref{eq_hi_vertical} to estimate the percentage of the column density contribution from the near and far side for each line-of-sight to solve the kinematic distance ambiguity.

\begin{table}
\caption{Coefficients of Eq.~\ref{eq_hi_vertical} taken from \citet{Dickey1990}.}
\centering
\label{table_hi_vertical}
\begin{tabular}{l c c}
\hline\hline
i & $n_i(0)$ (cm$^{-3}$) &$h_i$ (pc)\\
\hline
 1    & 0.395 &  90\\
 2   & 0.107 &  225\\
 3   & 0.064 &  403\\
\hline   
\end{tabular} 
\end{table}

\begin{figure}
\centering
\includegraphics[width=0.5\textwidth]{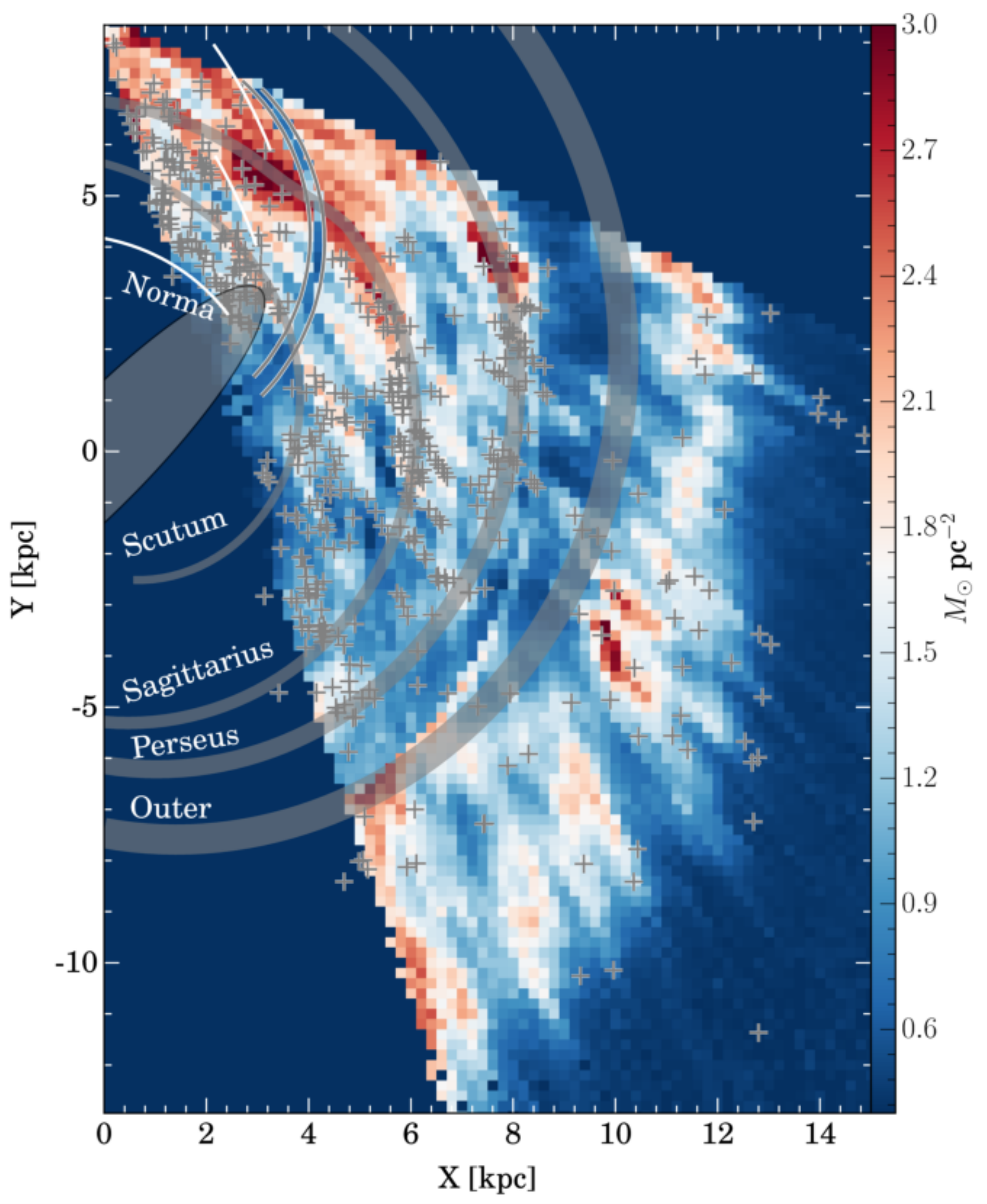}
\caption{Face-on view of \ion{H}{i} surface mass distribution in the survey area overlaid with spiral arms from \citet{Reid2016}, and \ion{H}{ii} regions from \citet{Anderson2014}. The Galactic center is at $\left[0,~0\right]$, and the Sun is at the top-left corner at a Galactocentric distance of 8.34~kpc \citep{Reid2014}. The gray lanes mark the Outer, Perseus, Sagittarius, and Scutum arms with widths from \citet{Reid2014}. The Local Spur, Aquila Spur, and Norm Arm are marked with white lines. The long bar \citep{Hammersley2000, Benjamin2005, Nishiyama2005, Benjamin2008, Cabrera2008} is marked with the shaded half-ellipse. The crosses mark the \ion{H}{ii} regions from \citet{Anderson2014} excluding the ones that fall into the tangent region (marked with two gray curves).}
 \label{fig_faceon}
\end{figure}

With the kinematic distances determined for each channel, we applied a 5$\sigma$ cut and converted the column density cube into a face-on mean surface density map, shown in Fig.~\ref{fig_faceon}. Comparing to the spiral arm model from \citet{Reid2016}, some of the atomic gas follows the spiral arms well, such as the Sagittarius and Perseus arms, but there is also much atomic gas in the inter-arm regions. Along the Sagittarius and Perseus arms, and in the very outer region beyond the Outer Arm in Fig.~\ref{fig_faceon}, the \ion{H}{ii} region distribution agrees with the atomic gas. However, the Outer Arm itself is not associated with much atomic gas in the face-on plot, although there is good agreement between the \ion{H}{I} emission and the Outer Arm in the $l-v$ diagram. This outer component of the atomic gas was also observed by \citet{Oort1958}, \citet{Nakanishi2003}, and \citet{Levine2006}. \citet{Nakanishi2003} found this emission structure agrees spatially with the Outer Arm discovered by \citet{Weaver1970}. The Perseus Arm and Outer Arm in Fig.~\ref{fig_faceon} are found to be distinct structures in the atomic gas distribution with a void of \ion{H}{i} emission and \ion{H}{ii} regions between the arms. The absolute value of the derived surface density represents a mean surface density along the $z$ direction, and is therefore lower than the results of \citet{Nakanishi2003} and \citet{Nakanishi2016}, which derived a surface density integrated along $z$.

We applied different Galactic models and rotation curves for the kinematic distance determination to create the face-on surface density maps shown in Fig.~\ref{app_faceon}. We find that the face-on density map depends only slightly on the assumed model for the kinematic distance. Compared to Fig.~\ref{fig_faceon} (Galactic parameters from \citealt{Reid2014}, rotational curve from \citealt{Persic1996}), the assumption of a uniform rotational curve (Fig.~\ref{app_faceon}, right) only lowers the gas surface density close to the bar region. Similar changes occur when applying the IAU Solar parameters ($R_0=8.5$~kpc, $\Theta_0=220$~km~s$^{-1}$) and the \citet{Brand1993} Galactic rotation model (Fig.~\ref{app_faceon}, left), together with slightly shifting the gas to higher distances.

\subsubsection{Mean spin temperature of the atomic gas}
With the \ion{H}{i} emission, we can obtain the line-of-sight mean spin temperature or the density-weighted harmonic mean spin temperature using \citep{Dickey2000}:
%-
\begin{equation}
\left<T_{\rm S}\right> = \frac{\int T_{\rm B}(v)\ dv} {\int 1-e^{-\tau(v)}\ dv}.
\label{eq_ts}
\end{equation}

\begin{figure}
\centering
\includegraphics[width=0.5\textwidth]{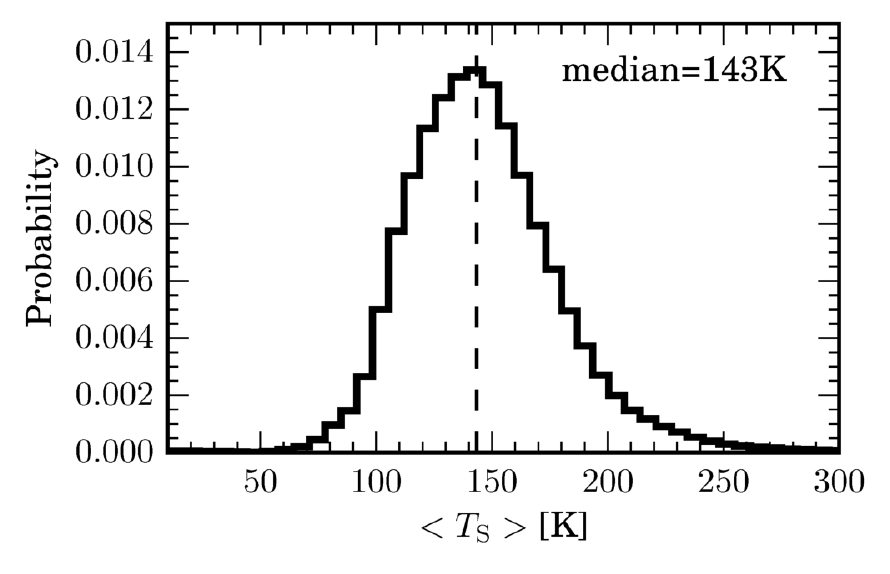}
\caption{Histograms of density-weighted harmonic mean spin temperature between --113 to 163~km~s$^{-1}$. The dashed vertical lines mark the median values of 143~K.}
 \label{fig_ts}
\end{figure}

As we show in Fig.~\ref{fig_ts}, the mean spin temperatures in the THOR survey area are between 50 and 300~K, with a median value of 143~K. We do not see any correlation between $\left<T_{\rm S}\right>$ and longitudes. Combining survey data from VGPS \citep{stil2006}, CGPS \citep{Taylor2003}, and SGPS \citep{McClure2005}, \citet{Dickey2009} studied the atomic gas in the outer disk of the Milky Way (outside the solar circle). They found the mean spin temperature to be $\sim$250--400~K, and it stays nearly constant with the Galactocentric radius to $\sim$25~kpc. Since the mean spin temperature $\left<T_{\rm S}\right>$ reveals the fraction of CNM in the total atomic gas (see Equation~8 in \citealt{Dickey2009}), the lower $\left<T_{\rm S}\right>$ we obtained indicates we observe a higher fraction of CNM in our survey area. Since \citet{Dickey2009} only considered the outer Milky Way, the differences in the $\left<T_{\rm S}\right>$ may indicate a higher fraction of CNM in the inner Milky Way.

\section{Discussion}
\label{sect_discuss}
We discuss the results of \ion{H}{i} overview presented in Sect.~\ref{sect_results} in this section.
\subsection{\ion{H}{i} gas mass}
\label{sect_hi_mass}
With the method described in Sect.~\ref{sect_hi_dist}, we estimate the optical depth corrected total \ion{H}{i} mass in our survey area to 4.7$\times10^8~M_\odot$ (using emission above 5$\sigma$). Combining $^{12}$CO(1--0) and $^{13}$CO(1--0) from the GRS and the Exeter FCRAO CO Survey, \citet{Roman-Duval2016} estimated an H$_2$ mass in the inner Galaxy (inside the solar circle) of 5.5$\times10^8~M_\odot$ (GRS: $18\degr\leq l \leq55.7\degr$, |$b\leq1\degr$|; Exeter: $55\degr\leq l \leq100\degr$, $-1.4\leq b\leq1.9\degr$). Within the solar circle, we estimate the total \ion{H}{i} mass to be 2.1$\times10^8~M_\odot$ ($18\degr\leq l \leq67\degr$, |$b|\leq1.25\degr$). Although the GRS+Exeter survey covers a larger area than THOR, the area they covered inside the solar circle is only $\sim4\%$ larger than THOR, so we can still compare these two masses. Therefore, in the inner Galaxy in this longitude range, the molecular component represents about 72\% of the total gas. This fraction also agrees with the molecular fraction of 50-60\% of total gas inside the solar circle estimated by \citet{Koda2016}, \citet{Nakanishi2016}, and \citet{Miville2017}. Considering the total \ion{H}{i} mass we derived is a lower limit, this percentage is likely an upper limit. 

If we assume the \ion{H}{i} emission to be optically thin, the total \ion{H}{i} mass within the whole survey area is estimated to be $3.6\times 10^{8}~M_\odot$. Comparing this with our optical-depth corrected estimate of 4.7$\times 10^{8}~M_\odot$, we find that the mass increases by $\sim$31\% percent with the optical depth correction. Considering that all of the $\tau$ spectra saturate in some channels, this $31\%$ is again a lower limit. We define the ratio between the mass after and before the optical depth correction as $R_{\rm \ion{H}{I}}$, so for the whole survey area $R_{\rm \ion{H}{I}}=1.31$. Assuming optically thin emission, the total mass of the \ion{H}{i} gas of the Milky Way within the Galactic radius of 30~kpc was estimated to be 7.2--8$\times10^{9}~M_\odot$ \citep{Kalberla2009, Nakanishi2016}. If we apply $R_{\rm \ion{H}{I}}=1.31$ to the whole Milky Way, the total \ion{H}{i} gas mass would be 9.4--10.5$\times10^{9}~M_\odot$. We discuss the uncertainties of the total \ion{H}{i} mass in Sect.~\ref{sect_mass_err}. 

As part of the pilot study of the THOR project, \citet{bihr2015} estimated the mass of the atomic hydrogen gas in the W43 region after applying the correction for optical depth and absorption against the diffuse continuum emission and derived an $R_{\rm \ion{H}{I}}=2.4$, which is higher than what we have derived for the entire survey. Considering W43 has the second largest integrated $\tau$ in the survey, it is reasonable that a larger than average $R_{\rm \ion{H}{I}}$ was measured here. The optical depth map (Fig.~\ref{fig_tau_map}) also shows higher $\tau$ in the inner longitude region ($l\lesssim 44\degr$) than in the outer longitude region.  

In contrast to this, \citet{Lee2015} found $R_{\rm \ion{H}{I}}\sim1.1$ by applying the optical depth correction pixel-by-pixel towards the atomic gas around the Perseus molecular cloud. Combining the 21~cm emission maps from the Galactic Arecibo L-band Feed Array Survey \citep[GALFA-HI][]{Peek2011, Peek2018} and absorption spectra from 21-{\it SPONGE} \citep{Murray2015, Murray2018b}, \citet{Murray2018a} found $1.0<R_{\rm \ion{H}{I}}<1.3$ towards high latitude local ISM ($|b|>15\degr$). Considering that we are observing with much higher angular resolution than GALFA, and multiple spiral arms along the line of sight are at the same velocity in the Galactic plane, it is reasonable that the $R_{\rm \ion{H}{I}}$ we derived is higher. On the other hand, studies toward nearby galaxies (M31, M33 and LMC) found $R_{\rm \ion{H}{I}}\sim1.3-1.34$ \citep{Braun2009, Braun2012}, which is in agreement with what we found.

\subsection{\ion{H}{i} gas distribution}
Ever since \citet{vandeHulst1954} and \citet{Oort1958} discovered the spiral structure of the atomic hydrogen gas in the Milky Way, considerable effort has been devoted to investigating the gas distribution and spiral arms in the Galaxy \citep[e.g.,][]{Kulkarni1982, Nakanishi2003, Levine2006, Kalberla2009, Nakanishi2016}. The structure outside of the Outer Arm in Fig.~\ref{fig_faceon} was also observed by \citet{Oort1958}, \citet{Nakanishi2003}, \citet{Levine2006}, and \citet{Nakanishi2016}. \citet{Nakanishi2003} fit this structure with the so-called Outer Arm with a pitch angle of $\sim7\degr$ in the polar coordinates (the x-axis is the azimuthal angle $\theta$ and the y-axis the Galactic radius $R$ in log scale). They found that this agrees with the Outer Arm found by \citet{Weaver1970}. Also, one of the arms fit by \citet{Levine2006} goes through the northern part of this structure (Y~$>0$, X~$>10$ in Fig.~\ref{fig_faceon}). \citet{Levine2006} claimed that the spiral arm model derived from the \ion{H}{ii} regions \citep{Morgan1953, Georgelin1976, Wainscoat1992} could not fit this structure, although Fig.~\ref{fig_faceon} shows that many \ion{H}{ii} regions are associated with this structure. The Outer Arm plotted in Fig.~\ref{fig_faceon} is extrapolated from the pitch angle fitted to parallax sources at greater longitudes \citep[$l>70\degr$,][]{Reid2016}. Since the pitch angle can vary by azimuth angle \citep{Honig2015}, the real Outer Arm in this region ($17\degr<l<67\degr$) could have a different pitch angle and be at a larger Galactocentric radius. On the other hand, the noncircular motions in the outer \ion{H}{i} disk \citep[e.g.,][]{Kuijken1994} could also affect the \ion{H}{i} distribution we derived in this region.

Another feature in the inner part of Fig.~\ref{fig_faceon} is that the region right below the end of the bar shows low surface density. Depending on the Galactic rotation model we used to determine the kinematic distances, there could even be a cavity in this region (all emission is below 5$\sigma$ and masked out, Fig.~\ref{app_faceon}). The region is located where the Galactic long bar ends \citep{Hammersley2000, Benjamin2005, Nishiyama2005, Benjamin2008, Cabrera2008}. The long bar introduces strong non-circular motions in the sources, and so the axisymmetric rotation curve does not apply on and near the bar \citep{Fux1999, Rodriguez2008, Reid2014}. Thus, the kinematic distances derived for the gas and most \ion{H}{ii} regions in this region may have large errors. Therefore, the low-level emission may not be real. The same low-level distribution is also seen in the CO map by \citet{Roman-Duval2009, Roman-Duval2016}, \ion{in H}{ii} regions. Considering all these distributions are based on kinematic distance (except for the distance of less than 10\% \ion{H}{ii} regions in this region are derived from parallax), we should consider the distribution of gas and \ion{H}{ii} regions inside Scutum Arm with caution.

\subsection{Atomic to molecular gas ratio}
\label{sect_hi_ratio}
Our optical depth correction discussed in the previous sections involves a lot of interpolation and averaging. We would like to avoid doing it when comparing the \ion{H}{i} and $^{13}$CO maps, since interpolation and averaging brings large uncertainties to each particular region. In the following discussion, we compare the \ion{H}{i} and $^{13}$CO emission directly. 

As we demonstrate in Fig.~\ref{fig_faceon}, high column density gas is concentrated along and above the Sagittarius Arm in the inner Galaxy. This is also seen by \citet{Nakanishi2016}. On the other hand, molecular clouds traced by CO are mainly distributed around and inside the Scutum Arm \citep{Roman-Duval2009, Nakanishi2016}. The molecular cloud fraction ($f_{\rm mol}$) map in \citet{Nakanishi2016} shows that $f_{\rm mol}$ is anti-correlated with Galactocentric distance. Compared to Fig.~\ref{fig_faceon}, inside the Sagittarius Arm, the molecular cloud fraction may be as high as $f_{\rm mol}>0.6$, while in the outer regions, $f_{\rm mol}$ quickly drops to almost zero \citep[see also, ][]{Miville2017}. However, they did not discuss how this ratio varies between inter-arm regions and spiral arm regions.

As the $l-v$ diagrams in Fig.~\ref{fig_hi_pv} and Fig.~\ref{fig_ratio_pv} show that the majority of the molecular gas is tightly associated with the spiral arms, while the atomic gas on the other hand is more widely distributed with significant material in the inter-arm regions. The histogram of $T_{\rm B}$(\ion{H}{i})/$T_{\rm B}$($^{13}$CO) ratio shows a bimodal distribution (Fig.~\ref{fig_ratio_hist}). If we assume both $^{13}$CO and \ion{H}{i} emissions are optically thin, the $T_{\rm B}$(\ion{H}{i})/$T_{\rm B}$($^{13}$CO) ratio bimodal distribution could be proxy of the atomic-to-molecular gas ratio. Therefore, we can interpret from Fig.~\ref{fig_ratio_hist} that on average, the atomic-to-molecular gas ratio may increase by approximately a factor of six from spiral arms to inter-arm regions.

\subsection{Uncertainties of the \ion{H}{I} optical depth $\tau$}
The uncertainties of the \ion{H}{i} optical depth are determined mainly by two factors, the brightness of the continuum source and the rms noise of the absorption spectra. The ratio of these two is the S/N ratio. As we mentioned in Sect.~\ref{sec_tau}, we selected sources with an S/N ratio $>$6 to ensure that the spectra have real absorption and the $\tau$ spectra is not saturated in all channels. Thus, the S/N ratio for the selected spectra ranges between six and 250, with a median value of 12, and the uncertainties associated with the rms noise are between 0.18 to 0.004 with a median value of 0.09, depending on the brightness of the continuum source. 

\begin{figure}
\centering
\includegraphics[width=0.45\textwidth]{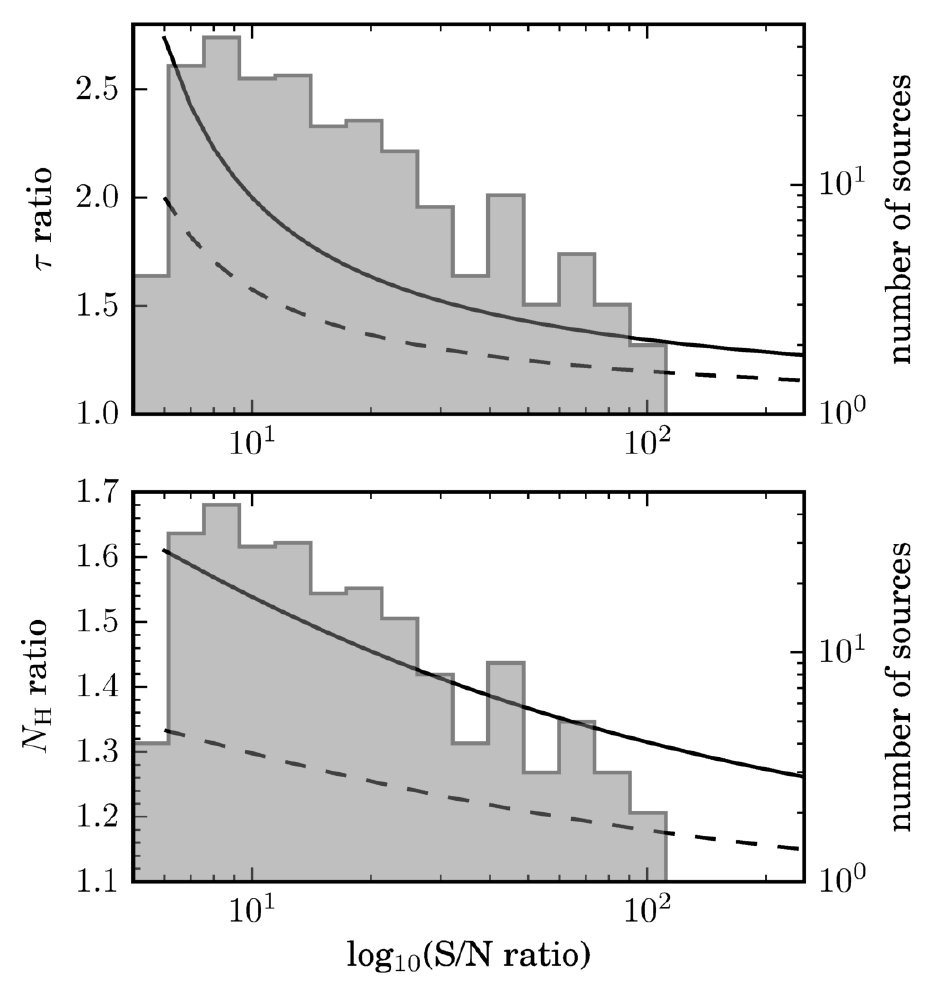}
\caption{Increase of the optical depth lower limit (top panel) and the column density (bottom panel) if the noise drops by 50\% (dashed line) and 70\% (solid line) plotted as a function of S/N ratio. The shaded histogram in each panel is the S/N ratio distribution of the continuum sources we used to extract the \ion{H}{i} absorption spectra and construct the optical depth map. }
 \label{fig_ratio}
\end{figure}

Another source of uncertainty for the optical depth $\tau$ is that the $\tau$ spectra are all saturated in some channels, as shown in Fig.~\ref{fig_spectra}. To test the $\tau$ limits, high sensitivity VLA follow-up observations were carried out towards three selected sources, G21.347--0.629, G29.956--0.018, and G31.388--0.384 (Rugel et al., in prep.). In the THOR survey, where each field was observed for five to six~minutes in on-source time, the optical depths of these three sources all saturate at $\tau\sim$2.7 to 3.1. In the follow-up observations, each continuum source was observed for significantly longer, and the noise dropped by $\sim50-70\%$. However, many channels in the $\tau$ spectra still saturate at $\tau\sim3.5-3.9$ if we smooth the data to the same angular and spectral resolution as the THOR C-configuration data. Depending on the S/N ratio of the continuum source, this $\sim50-70\%$ drop in the noise could increase the lower limit of the optical depth to $\sim1.2-2.6$ times the current value, and also increase the optical-depth corrected column density to $\sim 1.1-1.6$ times the current value (Fig.~\ref{fig_ratio}).

\subsection{Uncertainties of the \ion{H}{I} mass}
\label{sect_mass_err}
The uncertainties of the mass estimation for atomic gas originate from several factors: the optical depth, the distance, absorption against the diffuse continuum emission, and the \ion{H}{i} self absorption (HISA). As we discussed in the previous section, we could be underestimating the peak optical depth $\tau_{\rm limit}$ by 20 to 160\%, but the impact on the integrated optical depth is not clear. If we assume the noise levels of the $\tau$ spectra drop to 30\% of the values listed in Table.~\ref{table_tau}, with the same 228 sources to correct the optical depth, the total \ion{H}{I} mass increases by 9\%. However, as we mentioned in the previous section, many channels in the $\tau$ spectra still saturate in the high sensitivity follow-up observations, and this 20\% is still a lower limit.

The second main factor of uncertainty is the distance. With different Galactic parameters, we get different kinematic distances, which result in different mass estimates. With the rotation curve of \citet{Persic1996}, and the Galactic parameters from \citet{Reid2014}, we derived the total mass 4.7$\times10^8 M_\odot$. With the same Galactic parameters from \citet{Reid2014} but assuming a uniform rotational curve, we would get the same mass. If we take the IAU Solar parameters ($R_0=8.5$~kpc, $\Theta_0=220$~km~s$^{-1}$) and Galactic rotation model from \citet{Brand1993}, the total mass increases by 13\%. We are also aware that assuming axisymmetry and using the average vertical density distribution $n(z)$ of \ion{H}{i} to solve the kinematic distance ambiguity of the column density distribution is not ideal, but this is the best we can do without detailed modeling of the Galactic disk. Furthermore, \citet{Wenger2018} compared the kinematic distances with the parallax distances of 75 Galactic high-mass star-forming regions, and they found out that the kinematic distances we used in this paper (derived with the rotation curve of \citet{Persic1996} and the Galactic parameters from \citet{Reid2014}) have a median offset of 0.43~kpc, with a standard deviation of 1.24~kpc from parallax distances.

\citet{bihr2015} estimated the mass of the atomic hydrogen gas in W43 to be $6.6_{-1.8}\times10^6~M_\odot$ after applying the correction for optical depth and absorption against the diffuse continuum emission ($l=29.0-31.5\degr,\ |b|\leq 1\degr,\ v_{\rm LSR}=60-120$~km~s$^{-1}$). By integrating across the same velocity range, we derived a mass of 8.5$\times10^6~M_\odot$ within the same area with our distance determination method. The mean distance at $l=30.25\degr$ between 60 to 120~km~s$^{-1}$ from our method is 7.4~kpc, much larger than the distance of 5.5~kpc \citep{Zhang2014} adopted by \citet{bihr2015}. Taking the same distance 5.5~kpc results in a mass of $4.9\times10^6~M_\odot$. Further applying correction for the absorption against the diffuse continuum emission with the method described in \citet{bihr2015} increases the mass by 19\% to $5.7\times10^6~M_\odot$, which approximately agrees with what \citet{bihr2015} estimated, but is slightly smaller. Since \citet{bihr2015} use one single strong continuum source to correct the optical depth for the whole W43 region, they may overcorrect for the outer region in W43. The 19\% mass increase from applying a correction for the absorption against the diffuse continuum emission is an upper limit if we consider the whole survey area, since W43 is one of the most extreme star-forming complexes in the Milky Way and is considered to be a mini starburst region \citep{Nguyen2011, Beuther2012, Nguyen2013, Zhang2014}. Furthermore, to correct the diffuse continuum emission, we need information on its distance. It is reasonable to assume all diffuse continuum emission is in the background and correct it for a particular region, but we can not apply this to the whole survey. Thus, we do not apply any correction for the diffuse continuum emission.

The last factor is HISA, which we did not consider for the mass estimation. However, HISA was studied towards a Giant Molecular Filament (GMF) with the THOR data by Wang et al., submitted, and they showed, for this specific GMF, that the mass traced by HISA is only 2-4\% of the total \ion{H}{i} mass. Studies of \ion{H}{i} narrow self absorption (HINSA) show that the ratio between the column density traced by HINSA to the H$_2$ column density is $\sim 10^{-3}$ to $2\times 10^{-2}$ \citep{Li2003, Goldsmith2005, Zuo2018}. As we discussed in Sect.~\ref{sect_hi_mass}, the total H$_2$ mass in this part of the Galactic plane is also comparable to the \ion{H}{i} mass, the effect of HISA and HINSA to the \ion{H}{i} mass is negligible. In summary, we could still be underestimating the \ion{H}{i} mass by 20 to 40\%, even with the current optical depth correction. 

\section{Conclusions}
\label{sect_con}
In this paper, we describe the THOR data release 2, which includes all OH and RRL data, and an entire new \ion{H}{i} dataset from the THOR survey. In addition, a detailed analysis of the \ion{H}{i} data is presented. The main results can be summarized as follows:
\begin{enumerate}
\item While the \ion{H}{i} channel map shows clear filamentary substructures at negative velocities, the emission at positive velocities is more smeared-out. This is likely due to higher spatial and velocity crowding of structures at positive velocities. Both the $l-v$ diagram and the face-on view of the \ion{H}{i} emission show that some of the atomic gas follows the spiral arms well, such as the Sagittarius and Perseus Arm, but there is also much gas in the inter-arm regions. 
\item We produced a spectrally resolved \ion{H}{i} $\tau$ map from 228 absorption spectra. 
\item We corrected optical depth for the \ion{H}{i} emission with the $\tau$ map. The atomic gas column density we derived with optical depth correction is 38\% higher than the column density derived with optically thin assumption. We estimate the total \ion{H}{i} mass in the survey region to be 4.7$\times10^8~M_\odot$, 31\% higher than the mass derived with the optically thin assumption. If we apply this 31\% correction to the whole Milky Way, the total atomic gas mass would be 9.4--10.5$\times 10^9~M_\odot$. 
\item Considering that all the $\tau$ spectra are saturated in some channels and we did not apply the correction for the diffuse continuum emission, we could be underestimating the mass by an additional 20--40\%. Future higher sensitivity observations are needed to better constrain the optical depth.
\item We constructed a face-on view of the mean surface density of the atomic gas in the survey area.
\item We estimated the density-weighted harmonic mean spin temperature $\left<T_{\rm S}\right>$ integrated between --113 and 165~km~s$^{-1}$ with a median value of $\left<T_{\rm S}\right>\sim143$~K, about a factor of two lower than what was estimated in the outer disk of the Milky Way, which may indicate a higher fraction of CNM in the inner Milky Way.
\item The latitude averaged $T_{\rm B}$(\ion{H}{i})/$T_{\rm B}$($^{13}$CO) ratio distribution shows two peaks at $\sim$100 and $\sim$600, which may indicate that the atomic-to-molecular gas ratio can increase by a factor of six from spiral arms to inter-arm regions.
\end{enumerate}

The \ion{H}{i}, OH, RRL, and continuum data from the THOR survey together provide the community the basis for high-angular-resolution studies of the ISM in different phases.

\begin{acknowledgements}
The National Radio Astronomy Observatory is a facility of the National Science Foundation operated under cooperative agreement by Associated Universities, Inc. Y.W., H.B., S.B., and J.D.S. acknowledge support from the European Research Council under the Horizon 2020 Framework Program via the ERC Consolidator Grant CSF-648505. H.B., S.C.O.G., and R.S.K. acknowledge support from the Deutsche Forschungsgemeinschaft in the Collaborative Research Center (SFB 881) ``The Milky Way System'' (subproject B1, B2, B8). This work was carried out in part at the Jet Propulsion Laboratory which is operated for NASA by the California Institute of Technology. R.J.S. acknowledges an STFC Rutherford fellowship (grant ST/N00485X/1). N.R. acknowledges support from the Max Planck Society through the Max Planck India Partner Group grant. F.B. acknowledges funding from the European Union’s Horizon 2020 research and innovation program (grant agreement No 726384). This research made use of Astropy and affiliated packages, a community-developed core Python package for Astronomy \citep{astropy2018}, Python package {\it SciPy}\footnote{\url{https://www.scipy.org/}}, APLpy, an open-source plotting package for Python \citep{robitaille2012}, and software TOPCAT \citep{taylor2005}.

\end{acknowledgements}

\bibliographystyle{aa}
\bibliography{references.bib}

\begin{appendix}
\section{Optical depth measurements towards the selected sources}

\longtab[1]{
\begin{longtable}{l c c c c c r } 
\caption{\label{table_tau} Optical depth measurements towards the selected continuum sources.}\\
\hline\hline
Gal.ID& $T_{\rm cont}$   &  rms & $\tau_{\rm limit}$      &  $\int \tau(v) dv$ &$R_{\rm eff}$\footnotemark[1] & Note\footnotemark[2]\\
      & (K)                         & (K)   &                                      &  (km s$^{-1}$)   &\arcsec               & \\
\hline
\endfirsthead 
\caption{continued.}\\
\hline\hline
Gal.ID& $T_{\rm cont}$   &  rms & $\tau_{\rm limit}$      &  $\int \tau(v) dv$ &$R_{\rm eff}$\footnotemark[1] & Note\footnotemark[2]\\
      & (K)                         & (K)   &                                        &  (km s$^{-1}$) &\arcsec & \\
\hline
\endhead
\hline                     
\endfoot
  G14.490+0.021 & 117 & 17 & 0.8 & 30 & 22 & HII\\
  G15.035--0.677 & 4286 & 17 & 4.4 & 33 & 289 & HII\\
  G15.168+0.797 & 134 & 13 & 1.2 & 28 & 23 & \\
  G15.190--0.596 & 244 & 20 & 1.4 & 17 & 111 & HII\\
  G15.913+0.183 & 97 & 16 & 0.7 & 16 & 74 & SNR\_green\\
  G16.557+0.453 & 99 & 13 & 0.9 & 26 & 19 & \\
  G16.733--1.185 & 218 & 23 & 1.1 & 26 & 21 & Xray\\
  G16.784--1.058 & 162 & 16 & 1.2 & 23 & 40 & jet\\
  G16.945--0.074 & 114 & 12 & 1.1 & 59 & 30 & HII\\
  G17.910+0.372 & 244 & 11 & 2.0 & 53 & 21 & \\
  G18.092+1.167 & 145 & 20 & 0.9 & 15 & 22 & \\
  G18.106+0.186 & 179 & 12 & 1.6 & 65 & 24 & \\
  G18.148--0.283 & 462 & 15 & 2.3 & 40 & 99 & HII\\
  G18.270--0.289 & 110 & 15 & 0.9 & 16 & 184 & HII\\
  G18.303--0.390 & 542 & 13 & 2.6 & 43 & 32 & HII\\
  G18.696--0.401 & 97 & 14 & 0.8 & 40 & 20 & \\
  G18.755--0.497 & 83 & 10 & 1.0 & 54 & 19 & \\
  G18.761+0.287 & 106 & 17 & 0.7 & 26 & 252 & SNR\_green\\
  G18.886--0.508 & 120 & 14 & 1.1 & 26 & 145 & HII\\
  G19.075--0.287 & 281 & 15 & 1.8 & 45 & 223 & HII\\
  G19.432--0.824 & 79 & 13 & 0.7 & 25 & 20 & \\
  G19.492+0.135 & 223 & 14 & 1.7 & 76 & 100 & HII\\
  G19.610--0.235 & 889 & 15 & 3.0 & 80 & 112 & HII\\
  G19.621--0.701 & 70 & 10 & 0.9 & 25 & 19 & \\
  G19.679--0.131 & 153 & 17 & 1.1 & 55 & 45 & HII\\
  G20.080--0.136 & 145 & 13 & 1.3 & 67 & 33 & HII\\
  G20.252+0.991 & 119 & 14 & 1.0 & 24 & 23 & \\
  G20.594--0.130 & 109 & 13 & 1.0 & 66 & 18 & \\
  G20.750--0.090 & 89 & 13 & 0.8 & 35 & 175 & HII\\
  G20.923+0.213 & 89 & 15 & 0.7 & 21 & 23 & Xray\\
  G20.989+0.090 & 195 & 11 & 1.8 & 56 & 62 & HII\\
  G20.999--1.125 & 126 & 19 & 0.8 & 13 & 26 & PN\\
  G21.347--0.629 & 521 & 12 & 2.7 & 55 & 23 & Xray\\
  G21.503--0.884 & 1095 & 14 & 3.2 & 36 & 58 & SNR\_green\\
  G21.765--0.631 & 129 & 12 & 1.3 & 38 & 313 & SNR\_green\\
  G21.770+0.843 & 131 & 14 & 1.1 & 21 & 21 & \\
  G21.874+0.008 & 288 & 13 & 2.0 & 80 & 31 & HII\\
  G22.199--0.756 & 90 & 13 & 0.8 & 28 & 25 & \\
  G22.760--0.478 & 109 & 14 & 1.0 & 17 & 100 & HII\\
  G22.933--0.076 & 117 & 19 & 0.7 & 55 & 28 & jet;Xray\\
  G23.438--0.209 & 228 & 13 & 1.7 & 69 & 166 & HII\\
  G23.710+0.171 & 275 & 13 & 1.9 & 68 & 59 & HII\\
  G23.871--0.121 & 266 & 11 & 2.1 & 86 & 51 & HII\\
  G23.956+0.150 & 595 & 16 & 2.5 & 51 & 56 & HII\\
  G24.180+0.565 & 253 & 17 & 1.6 & 66 & 25 & \\
  G24.463+0.245 & 120 & 14 & 1.1 & 75 & 166 & HII\\
  G24.471+0.488 & 614 & 16 & 2.6 & 41 & 90 & HII\\
  G24.493--0.038 & 150 & 19 & 1.0 & 36 & 76 & HII\\
  G24.507--0.223 & 207 & 18 & 1.4 & 48 & 120 & HII\\
  G24.541+0.600 & 130 & 13 & 1.2 & 54 & 19 & \\
  G24.675--0.154 & 254 & 20 & 1.5 & 62 & 71 & HII\\
  G24.798+0.096 & 354 & 17 & 1.9 & 57 & 152 & HII\\
  G25.237--0.150 & 240 & 15 & 1.6 & 91 & 46 & jet\\
  G25.266--0.161 & 325 & 15 & 1.9 & 99 & 21 & Xray\\
  G25.361--1.102 & 186 & 22 & 1.0 & 15 & 23 & \\
  G25.395+0.033 & 190 & 18 & 1.2 & 73 & 48 & HII\\
  G25.397--0.141 & 1035 & 16 & 3.1 & 70 & 159 & HII\\
  G25.520+0.216 & 138 & 18 & 1.0 & 41 & 22 & HII\\
  G25.604--0.038 & 206 & 11 & 1.9 & 103 & 21 & \\
  G25.692+0.031 & 114 & 10 & 1.3 & 64 & 77 & HII\\
  G26.090--0.058 & 122 & 14 & 1.1 & 63 & 127 & HII\\
  G26.544+0.414 & 345 & 13 & 2.2 & 50 & 73 & HII\\
  G26.609--0.212 & 93 & 14 & 0.8 & 37 & 21 & HII\\
  G26.687--1.028 & 123 & 14 & 1.1 & 14 & 22 & \\
  G27.279+0.145 & 239 & 12 & 1.9 & 81 & 56 & HII\\
  G27.365+0.014 & 110 & 12 & 1.2 & 31 & 127 & SNR\_green\\
  G27.380+1.167 & 131 & 19 & 0.8 & 10 & 29 & jet\\
  G27.494+0.190 & 186 & 15 & 1.4 & 60 & 112 & HII\\
  G27.920+0.977 & 364 & 17 & 2.0 & 23 & 25 & \\
  G28.287--0.364 & 194 & 16 & 1.4 & 25 & 23 & HII\\
  G28.305--0.387 & 151 & 16 & 1.1 & 24 & 44 & HII\\
  G28.608+0.018 & 127 & 15 & 1.0 & 50 & 55 & HII\\
  G28.610--0.142 & 131 & 13 & 1.2 & 56 & 76 & SNR\_green\\
  G28.652+0.027 & 125 & 18 & 0.8 & 35 & 67 & HII\\
  G28.672--0.108 & 134 & 13 & 1.2 & 53 & 63 & SNR\_green\\
  G28.807+0.175 & 246 & 17 & 1.6 & 41 & 53 & HII\\
  G29.089+0.511 & 199 & 10 & 1.9 & 61 & 20 & \\
  G29.689--0.242 & 607 & 20 & 2.3 & 52 & 100 & SNR\_green\\
  G29.935--0.053 & 544 & 12 & 2.7 & 70 & 115 & HII\\
  G29.956--0.018 & 876 & 18 & 2.8 & 69 & 61 & HII\\
  G30.234--0.138 & 282 & 13 & 2.0 & 95 & 20 & HII\\
  G30.534+0.021 & 278 & 19 & 1.6 & 80 & 38 & HII\\
  G30.687--0.260 & 196 & 16 & 1.4 & 58 & 102 & \\
  G30.699--0.630 & 148 & 16 & 1.1 & 59 & 29 & jet\\
  G30.720--0.083 & 192 & 12 & 1.7 & 62 & 15 & HII\\
  G30.782--0.027 & 1669 & 27 & 3.0 & 80 & 226 & HII\\
  G31.070+0.050 & 152 & 13 & 1.3 & 88 & 38 & HII\\
  G31.242--0.110 & 244 & 18 & 1.5 & 86 & 31 & HII\\
  G31.279+0.064 & 140 & 17 & 1.0 & 36 & 25 & HII\\
  G31.388--0.384 & 768 & 11 & 3.1 & 110 & 21 & \\
  G31.394--0.259 & 142 & 17 & 1.0 & 48 & 56 & HII\\
  G31.412+0.307 & 401 & 24 & 1.7 & 48 & 27 & HII\\
  G31.869+0.064 & 410 & 19 & 1.9 & 50 & 193 & SNR\_green\\
  G32.151+0.133 & 265 & 17 & 1.6 & 33 & 35 & HII\\
  G32.265+1.168 & 265 & 29 & 1.1 & 38 & 23 & \\
  G32.272--0.226 & 153 & 15 & 1.2 & 48 & 28 & HII\\
  G32.363+0.934 & 162 & 16 & 1.3 & 52 & 30 & jet\\
  G32.389--0.403 & 208 & 13 & 1.7 & 50 & 20 & \\
  G32.798+0.191 & 687 & 15 & 2.7 & 71 & 40 & HII\\
  G32.835--0.730 & 81 & 13 & 0.7 & 9 & 21 & \\
  G32.928+0.607 & 133 & 11 & 1.4 & 60 & 29 & HII\\
  G33.133--0.093 & 92 & 12 & 0.9 & 42 & 28 & HII\\
  G33.143--0.066 & 98 & 12 & 1.0 & 58 & 22 & jet\\
  G33.417--0.004 & 88 & 13 & 0.8 & 30 & 55 & HII\\
  G33.498+0.194 & 484 & 9 & 2.8 & 103 & 20 & Xray\\
  G33.810--0.189 & 107 & 12 & 1.1 & 53 & 26 & HII\\
  G33.915+0.110 & 350 & 16 & 2.0 & 62 & 30 & HII\\
  G34.133+0.471 & 260 & 12 & 2.0 & 61 & 30 & HII\\
  G34.256+0.146 & 1343 & 17 & 3.2 & 66 & 83 & HII\\
  G34.568--0.630 & 78 & 11 & 0.8 & 18 & 153 & SNR\_green\\
  G34.588--0.238 & 107 & 13 & 1.0 & 14 & 127 & SNR\_green\\
  G35.053--0.518 & 92 & 14 & 0.8 & 26 & 85 & HII\\
  G35.139--0.762 & 105 & 16 & 0.8 & 11 & 30 & HII\\
  G35.467+0.139 & 130 & 16 & 1.0 & 37 & 23 & HII\\
  G35.574+0.068 & 193 & 16 & 1.4 & 55 & 27 & HII\\
  G35.947+0.379 & 93 & 11 & 1.0 & 50 & 22 & \\
  G36.056+0.357 & 246 & 11 & 2.0 & 82 & 26 & jet\\
  G36.060+0.994 & 91 & 14 & 0.8 & 25 & 18 & \\
  G36.516--0.970 & 95 & 12 & 1.0 & 27 & 20 & \\
  G36.551+0.002 & 476 & 10 & 2.7 & 94 & 21 & Xray\\
  G37.545--0.112 & 272 & 14 & 1.9 & 84 & 59 & HII\\
  G37.750--0.107 & 92 & 12 & 1.0 & 46 & 45 & HII\\
  G37.763--0.215 & 426 & 13 & 2.4 & 85 & 71 & HII\\
  G37.868--0.601 & 118 & 13 & 1.1 & 47 & 19 & HII\\
  G37.874--0.399 & 833 & 12 & 3.1 & 103 & 39 & HII\\
  G38.549+0.163 & 67 & 10 & 0.8 & 33 & 25 & HII\\
  G38.875+0.308 & 97 & 12 & 1.0 & 52 & 21 & HII\\
  G38.932--1.126 & 79 & 13 & 0.7 & 18 & 21 & \\
  G39.157+0.643 & 93 & 11 & 1.0 & 46 & 18 & \\
  G39.187--0.303 & 108 & 13 & 1.0 & 49 & 171 & SNR\_green\\
  G39.251--0.066 & 272 & 14 & 1.9 & 84 & 75 & HII\\
  G39.414--0.595 & 155 & 12 & 1.5 & 69 & 23 & \\
  G39.565--0.040 & 258 & 11 & 2.1 & 91 & 19 & \\
  G39.728--0.398 & 94 & 12 & 1.0 & 41 & 37 & HII\\
  G39.883--0.346 & 106 & 8 & 1.5 & 65 & 20 & HII\\
  G40.180--0.798 & 111 & 9 & 1.4 & 52 & 18 & \\
  G40.733+0.192 & 106 & 9 & 1.3 & 57 & 20 & \\
  G41.117--0.334 & 293 & 12 & 1.9 & 38 & 120 & SNR\_green\\
  G41.189+1.221 & 117 & 18 & 0.8 & 27 & 17 & \\
  G41.513--0.141 & 107 & 12 & 1.1 & 38 & 53 & HII\\
  G41.660+0.441 & 75 & 12 & 0.8 & 28 & 17 & \\
  G41.741+0.097 & 137 & 12 & 1.4 & 53 & 21 & HII\\
  G42.028--0.605 & 213 & 10 & 1.9 & 61 & 21 & \\
  G42.050+0.853 & 97 & 10 & 1.2 & 46 & 20 & \\
  G42.365+0.079 & 59 & 9 & 0.7 & 33 & 20 & jet\\
  G42.434--0.260 & 131 & 10 & 1.4 & 44 & 65 & HII\\
  G42.653+0.946 & 102 & 12 & 1.1 & 36 & 21 & \\
  G42.895+0.573 & 307 & 10 & 2.3 & 74 & 26 & Xray\\
  G43.171+0.007 & 2751 & 17 & 4.0 & 115 & 141 & HII\\
  G43.177--0.519 & 111 & 9 & 1.4 & 51 & 38 & HII\\
  G43.238--0.045 & 243 & 14 & 1.8 & 83 & 23 & HII\\
  G43.257--0.161 & 779 & 14 & 2.9 & 75 & 142 & SNR\_green;HII\\
  G43.738--0.620 & 168 & 10 & 1.8 & 54 & 22 & \\
  G43.890--0.783 & 152 & 11 & 1.5 & 41 & 40 & HII\\
  G43.921--0.479 & 120 & 14 & 1.1 & 45 & 28 & \\
  G44.006+0.959 & 69 & 11 & 0.8 & 27 & 18 & \\
  G44.492+1.045 & 110 & 13 & 1.0 & 32 & 21 & \\
  G44.649--0.795 & 92 & 9 & 1.2 & 46 & 25 & jet\\
  G45.067+0.138 & 122 & 11 & 1.3 & 32 & 35 & HII\\
  G45.122+0.132 & 738 & 10 & 3.2 & 49 & 60 & HII\\
  G45.454+0.059 & 1172 & 11 & 3.6 & 59 & 133 & HII\\
  G45.478+0.130 & 544 & 12 & 2.7 & 55 & 63 & HII\\
  G45.519--0.870 & 117 & 9 & 1.5 & 47 & 22 & \\
  G45.751+0.198 & 111 & 9 & 1.5 & 52 & 18 & \\
  G45.823--0.284 & 103 & 10 & 1.2 & 32 & 48 & HII\\
  G46.260--0.851 & 78 & 9 & 1.1 & 37 & 19 & \\
  G48.241--0.968 & 272 & 9 & 2.3 & 50 & 19 & \\
  G48.545--0.003 & 79 & 10 & 1.0 & 37 & 45 & HII\\
  G48.610+0.027 & 126 & 13 & 1.1 & 43 & 97 & HII\\
  G48.930--0.280 & 494 & 17 & 2.3 & 39 & 120 & HII\\
  G48.983--0.299 & 201 & 15 & 1.5 & 41 & 74 & HII\\
  G49.079--0.374 & 350 & 17 & 1.9 & 41 & 87 & HII\\
  G49.206--0.342 & 943 & 20 & 2.8 & 44 & 107 & HII\\
  G49.210--0.963 & 376 & 14 & 2.2 & 44 & 30 & jet\\
  G49.370--0.302 & 1428 & 26 & 2.9 & 54 & 144 & HII\\
  G49.477--0.328 & 465 & 16 & 2.3 & 43 & 32 & HII\\
  G49.488--0.380 & 3600 & 27 & 3.8 & 61 & 163 & HII\\
  G49.553--0.330 & 74 & 11 & 0.8 & 23 & 32 & HII\\
  G49.586--0.385 & 365 & 16 & 2.0 & 37 & 66 & HII\\
  G49.594--0.449 & 89 & 15 & 0.7 & 13 & 50 & HII\\
  G50.234+0.327 & 212 & 13 & 1.7 & 61 & 32 & jet\\
  G50.285--0.392 & 100 & 12 & 1.0 & 38 & 31 & HII\\
  G50.626--0.031 & 206 & 11 & 1.8 & 48 & 19 & \\
  G50.948+0.847 & 114 & 12 & 1.1 & 46 & 21 & \\
  G50.970+0.890 & 85 & 12 & 0.9 & 40 & 20 & \\
  G51.364--0.014 & 107 & 10 & 1.2 & 29 & 104 & SNR\_anderson\\
  G52.099+1.042 & 206 & 12 & 1.7 & 34 & 26 & HII\\
  G52.236+0.742 & 71 & 10 & 0.9 & 24 & 103 & HII\\
  G52.753+0.334 & 168 & 11 & 1.6 & 46 & 27 & HII\\
  G53.185+0.161 & 144 & 12 & 1.4 & 46 & 77 & HII\\
  G54.078--0.813 & 87 & 9 & 1.2 & 37 & 22 & \\
  G54.096--0.054 & 68 & 8 & 1.0 & 28 & 91 & HII\\
  G54.102+0.100 & 89 & 8 & 1.3 & 48 & 18 & \\
  G54.170--0.009 & 152 & 11 & 1.5 & 44 & 20 & \\
  G56.023+0.519 & 68 & 8 & 1.1 & 29 & 18 & \\
  G56.082+0.105 & 154 & 7 & 1.9 & 42 & 20 & Xray\\
  G56.348+0.394 & 149 & 9 & 1.7 & 52 & 23 & \\
  G56.616+0.170 & 56 & 8 & 0.8 & 23 & 19 & \\
  G57.547--0.272 & 133 & 11 & 1.4 & 37 & 54 & HII\\
  G57.701--0.456 & 69 & 6 & 1.3 & 41 & 17 & \\
  G58.189+0.931 & 62 & 9 & 0.8 & 23 & 19 & \\
  G58.966+0.220 & 195 & 9 & 2.0 & 62 & 33 & jet\\
  G59.483--0.881 & 156 & 9 & 1.8 & 46 & 17 & \\
  G59.672--0.206 & 51 & 7 & 0.8 & 27 & 17 & \\
  G59.801+0.231 & 78 & 8 & 1.1 & 30 & 71 & HII\\
  G60.804--0.632 & 74 & 10 & 0.9 & 25 & 44 & jet\\
  G60.883--0.130 & 133 & 11 & 1.4 & 19 & 37 & HII\\
  G61.050--0.880 & 79 & 8 & 1.3 & 31 & 21 & \\
  G61.475+0.092 & 1442 & 13 & 3.6 & 34 & 73 & HII\\
  G61.619+0.356 & 103 & 7 & 1.6 & 43 & 19 & \\
  G62.078+0.609 & 173 & 8 & 2.0 & 48 & 20 & \\
  G62.272+0.555 & 78 & 7 & 1.3 & 39 & 17 & \\
  G62.366--0.956 & 570 & 13 & 2.7 & 36 & 35 & jet\\
  G63.002+0.817 & 294 & 8 & 2.5 & 41 & 19 & \\
  G63.169+0.457 & 106 & 8 & 1.4 & 27 & 123 & HII\\
  G63.566+0.277 & 95 & 8 & 1.4 & 31 & 18 & \\
  G64.019--0.846 & 113 & 9 & 1.4 & 19 & 18 & \\
  G64.131--0.472 & 191 & 9 & 2.0 & 23 & 62 & HII\\
  G64.382--0.757 & 114 & 8 & 1.6 & 25 & 18 & \\
  G64.567+0.107 & 163 & 11 & 1.6 & 47 & 24 & jet\\
  G64.854--0.814 & 73 & 10 & 0.9 & 16 & 21 & \\
  G65.002--0.675 & 106 & 9 & 1.3 & 21 & 19 & \\
  G65.307--0.214 & 327 & 9 & 2.5 & 38 & 19 & \\
  G65.321+0.232 & 133 & 11 & 1.4 & 33 & 22 & \\
  G65.594+0.911 & 113 & 9 & 1.4 & 33 & 19 & \\
  G66.997--1.026 & 84 & 8 & 1.2 & 24 & 18 & \\
  G67.051+0.942 & 85 & 9 & 1.1 & 49 & 18 & \\
  G67.170+0.127 & 118 & 9 & 1.5 & 31 & 19 & \\
\end{longtable}
\footnotetext[1]{Effective radius $R_{\rm eff}$ was calculated from the area of the source assuming a circular shape.}
\footnotetext[2]{The physical nature of the continuum sources are taken from \citet{Wang2018}. ``HII'': sources associated with \ion{H}{ii} regions from \citet{Anderson2014}; ``SNR\_green'': sources associated with SNRs from \citet{Green2014}; ``SNR\_anderson'': sources associated with SNR candidates from \citet{anderson2017}; ``PN:'' sources classified as planetary nebula; ``Xray'': sources associated with X-ray sources;``jets'': sources classified as extragalactic jets candidates. The sources without a note in this column are most likely to be extragalactic origin.} 
}

\section{Face-on surface density maps of the atomic gas}

\begin{figure*}
\centering
\includegraphics[width=0.49\textwidth]{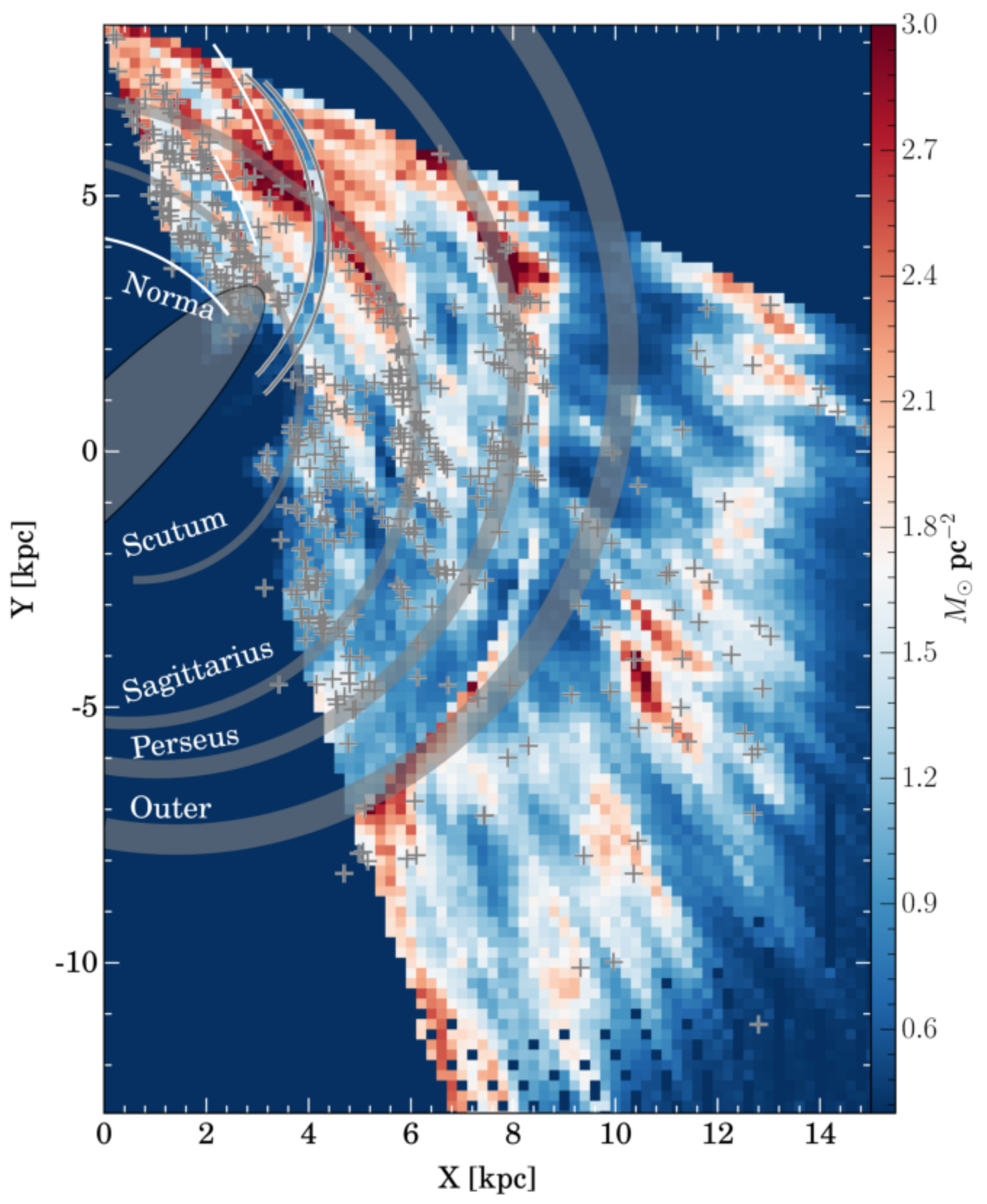}
\includegraphics[width=0.49\textwidth]{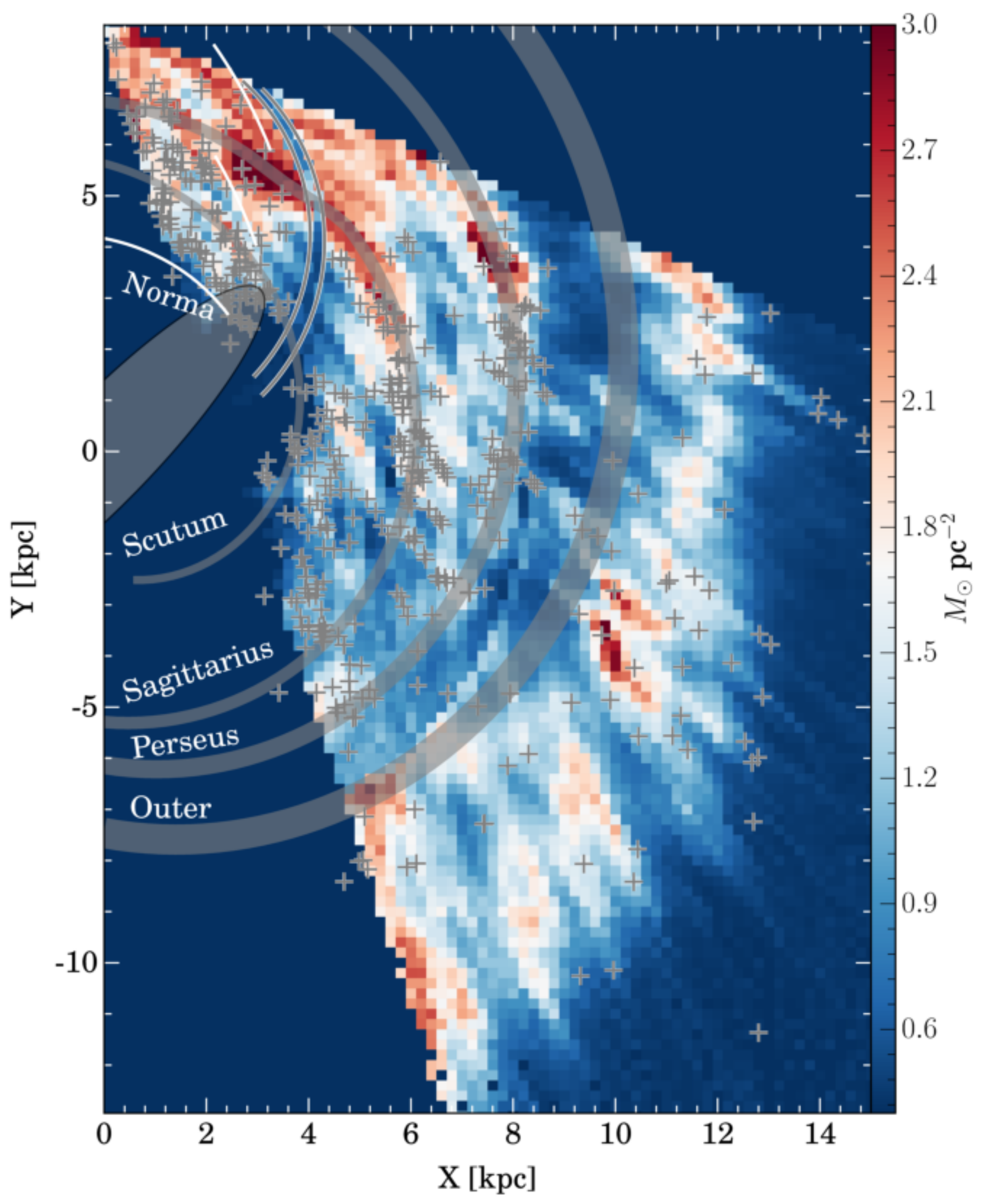}
\caption{Face-on view of the \ion{H}{i} surface mass distribution produced with different kinematic distance determination methods. Left panel: Kinematic distances were determined with IAU Solar parameters ($R_0=8.5$~kpc, $\Theta_0=220$~km~s$^{-1}$) and Galactic rotation model from \citet{Brand1993}. Right panel: Kinematic distances were determined with the Galactic parameters from \citet{Reid2014} with a uniform Galactic rotation curve. The rest of the symbols are the same as Fig.~\ref{fig_faceon}}
 \label{app_faceon}
\end{figure*}

\end{appendix}

\end{document}